\documentclass[aip,onecolumn,10pt]{revtex4}%
\usepackage{amsmath}
\usepackage{amsfonts}
\usepackage{amssymb}
\usepackage{graphics}
\usepackage{graphicx}
\providecommand{\U}[1]{\protect\rule{.1in}{.1in}}

\begin{document}
	\title{Multi-Path Interferometric Josephson Directional Amplifier for Qubit Readout}
	\author{Baleegh Abdo}
	\email{babdo@us.ibm.com}
	\author{Nick Bronn}
	\author{Oblesh Jinka}
	\author{Salvatore Olivadese}
	\author{Markus Brink}
	\author{Jerry M. Chow}
	\affiliation{IBM T. J. Watson Research Center, Yorktown Heights, New York 10598, USA.}
	\date{\today}

\begin{abstract}
We realize and characterize a quantum-limited, directional Josephson amplifier suitable for qubit readout. The device consists of two nondegenerate, three-wave-mixing amplifiers that are coupled together in an interferometric scheme, embedded in a printed circuit board. Nonreciprocity is generated by applying a phase gradient between the same-frequency pumps feeding the device, which plays the role of the magnetic field in a Faraday medium. Directional amplification and reflection-gain elimination are induced via wave interference between multiple paths in the system. We measure and discuss the main figures of merit of the device and show that the experimental results are in good agreement with theory. An improved version of this directional amplifier is expected to eliminate the need for bulky, off-chip isolation stages that generally separate quantum systems and preamplifiers in high-fidelity, quantum-nondemolition measurement setups.   
\end{abstract}

\maketitle

\newpage

%\pacs{84.30.Le, 42.25.Hz, 85.25.Cp, 78.20.Ls} 

%\noindent{\it Keywords}: nonreciprocal devices, directional amplifiers, Josephson amplifiers, quantum-limited amplification, qubit readout, quantum measurements 

\section{Introduction}
Performing single-shot, high-fidelity, quantum nondemolition (QND) measurements of quantum systems is crucial for running sophisticated quantum processing operations such as, quantum feedback \cite{JPADicarloReset,RabiVijay}, quantum error detection and correction \cite{FeedbackJPC}, and real-time tracking of quantum-state evolution \cite{FluxoniumJumps,JPCparity}. Such measurements are enabled using detection schemes that do not destroy the quantum state, have little to no classical backaction on the quantum system, preserve the observable being measured, and generate large signal to noise ratio (SNR) outcomes. 

A successful, commonly-used scheme for executing such measurements with superconducting qubits combines two measurement techniques, (1) dispersive readout to perform QND measurements of the qubit state \cite{QED,QEDcooperBox} and (2) use of a high-gain, quantum-limited output chain to significantly enhance the fidelity of the measurement \cite{QuantumJumps,QubitJPC}, that is enabled by amplifying the weak microwave signals carrying the qubit state information, while adding the minimum amount of noise required by quantum mechanics \cite{Caves,NoiseAmplReview}. Prominent components in such high-gain, low-noise output chains are quantum-limited (QL) Josephson amplifiers, such as Josephson parametric amplifiers (JPAs) \cite{JBA,VijayJBAreview,JBAMichael,CastellanosNat,Yamamoto} and Josephson parametric converters (JPCs) \cite{JPCnaturePhys,JPCnature,JPCreview} . To reduce quantum information loss and achieve high SNR of the output chain, these amplifiers are integrated as close as possible to the output of the measured quantum system. However, since these amplifiers work in reflection, nonreciprocal devices, such as circulators and isolators \cite{Pozar}, are needed to separate the input from output. These devices are also needed in order to protect the quantum system from reflected amplified signals and noise and from strong microwave pump tones driving these amplifiers \cite{VijayJBAreview}. Typically, to provide sufficient protection, high-fidelity measurement schemes employ $2-4$ cryogenic isolation components between the quantum system and the QL amplifier \cite{QuantumJumps,QubitJPC}. The addition of these cryogenic, nonreciprocal devices \cite{quinstar} to the measurement scheme is disadvantageous because they (1) are relatively large, heavy, and lack on-chip integration, all of which severely hinder scalability, (2) employ strong magnetic fields and materials that can negatively affect superconducting circuits, and (3) introduce insertion loss, causing about $50\%$-$70\%$ of the quantum signal to be lost in the intermediate stages of the isolation chain prior to reaching the QL amplifier.    

To mitigate these unwanted effects and possibly eliminate the need for intermediate circulators and isolators, several low-noise directional amplifiers have recently been proposed and developed, such as the Superconducting Low-inductance Undulatory Galvanometer (SLUG) amplifier \cite{SLUGThy,SLUGExp,HFslug}, the Josephson Traveling-Wave Parametric Amplifiers (JTWPA)\cite{TWPIrfan,JTWPA,TWPAthreewavemix}, the Kinetic Inductance Traveling-Wave Parametric Amplifiers (KITWPA) \cite{KIT}, and the multipump Josephson directional amplification schemes based on parametric interaction between three-modes residing in the same nonlinear device \cite{ReconfJJCircAmpl,NRAumentado1,NRAumentado2}. While all these directional amplifiers are viable and strong candidates for use in scalable quantum architectures, they each have shortcomings, which leave the door open for the development of new alternatives. For example, the SLUG dissipates energy on chip and can have out-of-band backaction. In addition, the JTWPA and KITWPA, due to their very broad bandwidth $2-5$ $\operatorname{GHz}$, are quite sensitive to impedance mismatches at their input and output circuitry as well as within the device itself, which can lead to ripples in the gain curve and undesired backaction on the quantum system. Also, the traveling-wave amplifier can generate, through its wave-mixing amplification process, undesired out-of-band harmonics due to its strong nonlinearity, which are difficult to suppress. Whereas the multipump Josephson directional amplifiers of Refs. \cite{ReconfJJCircAmpl,NRAumentado2} require the application of three microwave pump tones for their operation and have more degrees of freedom to control and stabilize, compared to the JTWPA and KITWPA for example, which can be unfavorable in scalable quantum architectures. 

In this work, we realize and characterize a two-port, integrated-circuit, Josephson directional amplifier suitable for qubit readout. The device is based upon the directional amplification scheme introduced and successfully demonstrated in Ref. \cite{JDA}. The scheme induces directional amplification and eliminates reflections by utilizing three key physical processes: (1) induced nonreciprocal phase shifts in nondegenerate, three-wave-mixing devices (i.e., JPCs), (2) signal build-up using a self-loop and partial attenuation of an internal mode of the system (i.e., via routing a portion of the idler power of the JPC towards auxiliary ports), and (3) generation of constructive/destructive wave-interference between multiple paths in the device. In order to distinguish it from other directional parametric amplifiers \cite{ReconfJJCircAmpl,NRAumentado2}, we call it the Multi-Path Interferometric Josephson Directional Amplifier (MPIJDA), which emphasizes its extrinsic interferometric mechanism.

The main properties of the MPIJDA scheme can be summarized as follows: it (1) amplifies signals propagating in the forward direction, i.e., input to output, by about $20$ dB, (2) amplifies signals propagating in the backward direction, i.e., output to input, by less than $2.5$ dB, (3) suppresses reflection gain on the input and output ports, (4) preserves the signal frequency across the device ports (i.e., the input and output signals have the same frequency), (5) amplifies both quadratures of the microwave field (i.e., functions as a phase-preserving amplifier), (6) adds noise of about a half input photon to the processed signal (i.e., operates near the quantum limit), (7) requires, in principle, one pump drive for its operation, (8) does not generate undesired out-of-band harmonics, (9) does not dissipate energy on chip, i.e., the Josephson junctions in the circuit stay in the zero-voltage state, and (10) does not leak power from the internal modes of the device and the pump drives to the input and output ports.

\begin{figure*}
	[tb]
	\begin{center}
		\includegraphics[
		width=0.8\columnwidth 
		]%
		{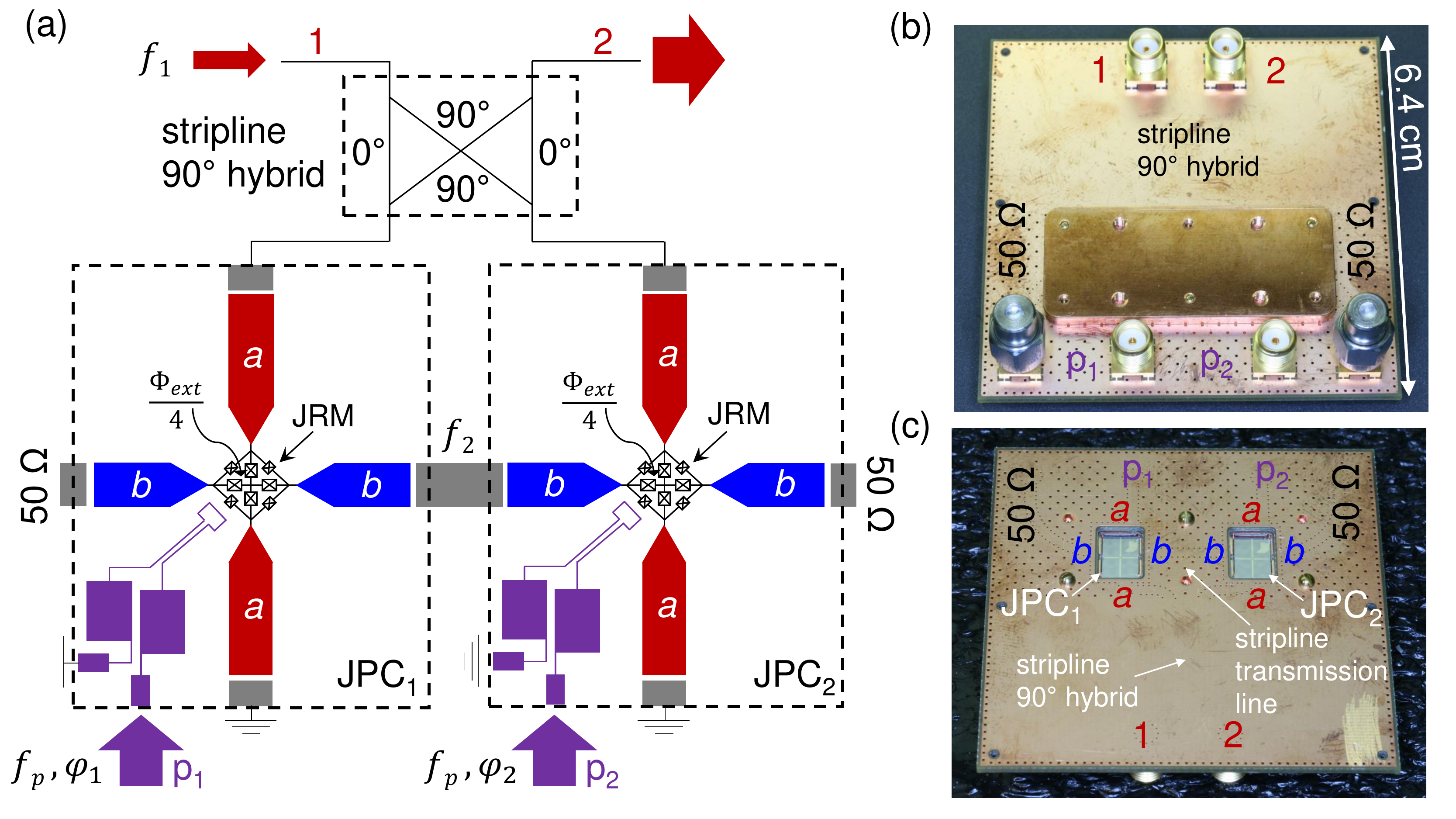}
		\caption{(a) Circuit diagram of the Multi-Path Interferometric Josephson Directional Amplifier (MPIJDA). The device consists of two nominally identical, hybridless Josephson Parametric Converters (JPCs) coupled through their ports \textit{a} and \textit{b} via a $90^{\circ}$ hybrid and a short transmission line, respectively. The diagram displays a cartoon of the center conductor of the microstrip JPC chips realized on $350$ $\mu$$\operatorname{m}$ thick silicon substrates. The two orthogonal resonators \textit{a} and \textit{b} intersect at a Josephson Ring Modulators (JRM) located at an rf-current anti-node of their respective fundamental, half-wavelength resonance modes. The JRM is a dispersive, three-wave-mixing element. It consists of four nominally identical Josephson junctions (JJs) arranged in a Wheatstone bridge configuration. The inner, four large JJs form a linear inductive shunting for the JRM, making it tunable with applied external flux \cite{Roch}. Both resonators \textit{a} and \textit{b} have two feedlines that are symmetrically coupled to either end via gap capacitances. One feedline of resonator \textit{a} is connected to the $90^{\circ}$ hybrid; the other is short circuited to ground. In the case of resonator \textit{b}, one feedline is connected to a $50$ Ohm termination; the other is part of a common short transmission line between the two JPCs. The pump drives, which parametrically amplify tones that lie within the bandwidths of resonators \textit{a} and \textit{b}, are fed to the JPCs through two separate feedlines that couple to the JRMs directly. The pump feedlines include sections of on-chip wide (capacitive) and narrow (inductive) lines, which form a filtering element inhibiting power leakage between the resonators and the pump ports. (b) and (c) exhibit photos of the top and bottom view of an integrated-circuit implementation of the MPIJDA, respectively. The bottom view photo does not show the copper cover that shields the JPCs chips or the two small, external superconducting magnetic coils attached to it that flux bias the two JRMs.              
		}
		\label{Device}
	\end{center}
\end{figure*}

\begin{figure*}
	[tb]
	\begin{center}
		\includegraphics[
		width=0.8\columnwidth 
		]%
		{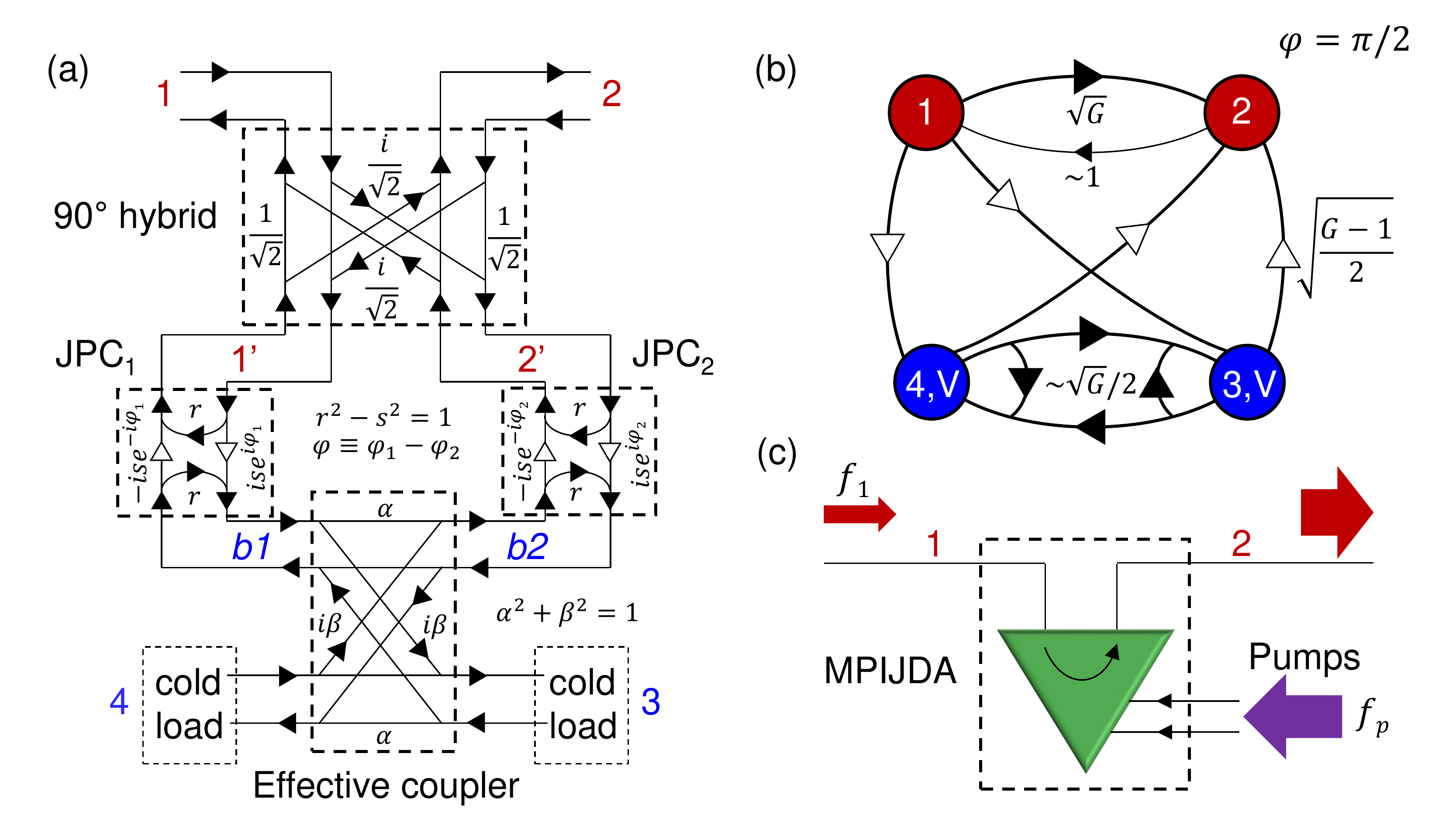}
		\caption{(a) Signal flow graph for the MPIJDA. Adapted from Ref. \cite{JDA} to match the content and symbols of the present paper. It consists of two JPCs coupled together through their signal ($1^{\prime}$, $2^{\prime}$) and idler (b1, b2) ports via two symmetric couplers. The first is a $90^{\circ}$ hybrid, which connects ports 1 (input) and 2 (output) of the MPIJDA to the signal ports ($1^{\prime}$, $2^{\prime}$) of the two JPCs. The second is a fictitious one, which models the attenuation present in the idler channel, due to the $50$ Ohm cold termination connected to one feedline of each \textit{b} resonator and the insertion loss of the normal-metal transmission line coupling the two stages. The latter coupler has generic, real coefficients $\alpha$ and $\beta$, which satisfy the condition ${\alpha}^2+{\beta}^2=1$. The transmission/reflection coefficients indicated on the graph correspond to on-resonance signals $f_1=f_a$, $f_2=f_b$. The transmitted signals between ports \textit{a} and \textit{b} of the JPC undergo frequency conversion. The unfilled arrows in the JPC flow graph, connecting ports \textit{a} and \textit{b}, represent phase conjugation. Directional amplification can be generated in the MPIJDA by applying a phase gradient $\varphi$ between the pump tones feeding the two JPCs. Maximum forward gain from input to output can be obtained for $\varphi=\pi/2$. (b) A graphical representation of the scattering parameters of the device for $\varphi=\pi/2$. Ports 1 and 2 are the input and output, respectively, while 3 and 4 are auxiliary ports inputting vacuum noise into the device (V). For clarity, scattering parameters with near zero amplitude are omitted. (c) A circuit symbol for the MPIJDA. 
		}
		\label{SignalFlow}
	\end{center}
\end{figure*}

\begin{figure}
	[tb]
	\begin{center}
		\includegraphics[
		width=0.8\columnwidth 
		]%
		{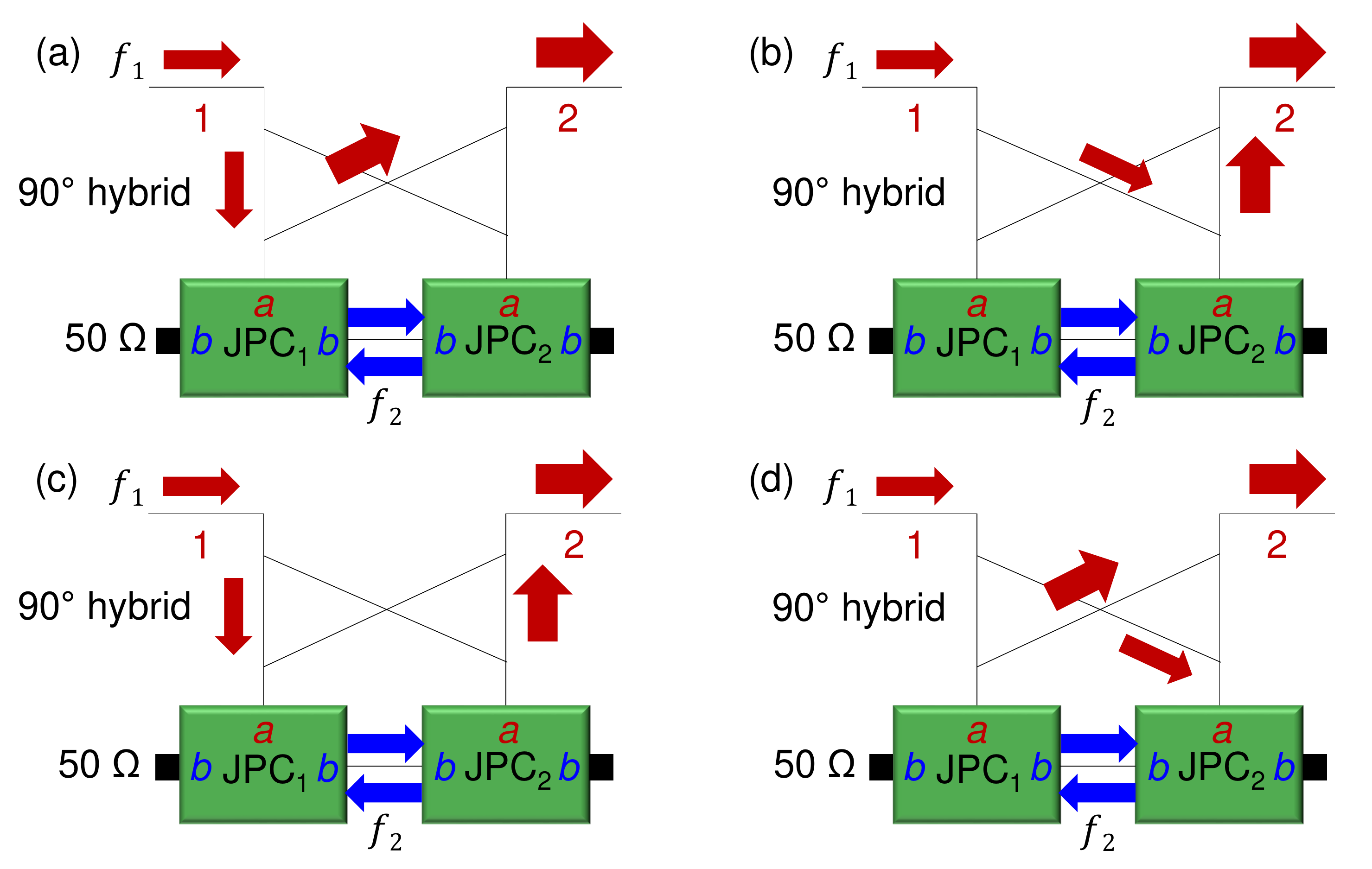}
		\caption{Illustration of the four principal wave-interference paths that contribute to the transmitted signal in the forward direction (i.e., from port $1$ to port $2$). A large amplified (attenuated), transmitted signal results if the wave interference between these principal paths is constructive (destructive). (a) The signal reflects off $\rm{JPC_1}$ port. (b) The signal reflects off $\rm{JPC_2}$ port. (c) The signal enters $\rm{JPC_1}$ and exits through $\rm{JPC_2}$. (d) The signal enters $\rm{JPC_2}$ and exists through $\rm{JPC_1}$. Similar principal paths can be constructed for transmitted signals in the opposite direction (i.e., from port $2$ to port $1$), or reflected signals [i.e., from port $1$ ($2$) back to port $1$ ($2$)].   
		}
		\label{MultiPathInterference}
	\end{center}
\end{figure}

\section{The device}  

The basic building block of the device is the JPC, which is a nondegenerate, Josephson parametric amplifier capable of performing  quantum-limited, phase-preserving amplification of microwave signals \cite{JPCnature,JPCreview}. The JPC [schematically shown in Fig. \ref{Device} (a)] consists of a Josephson Ring Modulator (JRM), which serves as a dispersive nonlinear medium, strongly coupled to two nondegenerate, superconducting resonators, denoted \textit{a} and \textit{b}. The JRM consists of four Josephson junctions arranged in a Wheatstone bridge configuration. The four large JJs inside the JRM form a linear inductive shunting for the JRM junctions, whose primary role is making the resonance frequencies of the JPC tunable with applied external flux \cite{Roch}. In the present realization, the resonators \textit{a} and \textit{b} are half-wave, microstrip resonators having fundamental resonance frequencies $f_{a,b}=\omega_{a,b}/2\pi$ and linear bandwidths $\gamma_{a,b}/2\pi\cong40$ $\operatorname{MHz}$. The two ends of each resonator are capacitively coupled to $50$ Ohm feedlines with equal gap capacitances \cite{Jamp}. This capacitive coupling to the feedlines sets the linear bandwidths of the JPC resonators. Each pair of feedlines coupled to the resonators form a single port for the JPC (hence, the JPC has two ports). When operated in the amplification mode, the JPC is driven by a nonresonant, common drive denoted as pump (P), whose frequency is the sum $f_p=f_a+f_b$ \cite{JPCreview,Jamp}. In this mode, weak signal (S) and idler (I) tones at frequencies $f_1=\omega_1/2\pi$ and $f_2=\omega_2/2\pi$, which lie within the dynamical bandwidth $B/2\pi$ of the device and satisfy $f_p=f_1+f_2$, get amplified equally in reflection. 

The pump drive is fed to the JPC using a separate physical feedline from those of resonators \textit{a} and \textit{b} \cite{hybridLessJPC}. The addition of this separate feedline allows us to do away with the off-chip, bulky, wideband, $180^{\circ}$ hybrids that are used when the resonators feedlines are employed for the purpose of addressing the JPC differential modes \textit{a},\textit{b} and injecting the pump drive to the JPC \cite{JPCreview}. The on-chip portion of the pump feedline [as shown in the circuit diagram in Fig. \ref{Device} (a)] consists of two narrow, coupled striplines that are shorted near the JRM. As demonstrated in Ref. \cite{hybridLessJPC}, such configuration enables the pump drive to differentially excite two adjacent nodes of the JRM, and thereby induce a common mode of the ring. The remainder of the on-chip pump line consists of two large-area capacitive pads and short inductive sections, which serve as a basic filtering element that suppresses power leakage between the resonators and the pump line. It is important to note that these hybridless JPC chips have the same area as those that do not incorporate the on-chip lines. Also, the on-chip pump lines can be used to flux-bias the JRM by applying dc-currents through them \cite{hybridLessJPC}. In this work, in order to avoid using bias tees for feeding the dc-currents and pumps to the same feedlines, we flux bias the JRMs using two small, external, superconducting magnetic coils attached to the device package. 

As shown in Fig. \ref{Device} (a), the MPIJDA is formed by coupling one feedline of the signal and idler resonators of two nominally identical JPCs via a $90^{\circ}$ hybrid \cite{Pozar} and a short transmission line, whose electrical length is about a wavelength at $f_2\approx10$ $\operatorname{GHz}$, respectively. The other feedline of the signal and idler resonators is shorted to ground and terminated by a $50$ Ohm load, respectively. The two ports of the MPIJDA, denoted as $1$ (input) and $2$ (output), are formed by the $90^{\circ}$ hybrid ports that are not directly connected to the signal feedlines of the JPCs. Figure \ref{Device} (b) and (c) show a top and bottom view of the MPIJDA realized in this work. The two JPC chips are integrated into a copper printed circuit board (PCB), whose dielectric substrate is FR4 with a relative dielectric constant of $\varepsilon_d=3.65$. The $90^{\circ}$ hybrid and shared transmission line coupling the signal and idler resonators are realized in the PCB using a stripline geometry. The other idler feedlines, which are not connected through the shared PCB transmission line, are terminated by external cryogenic $50$ Ohm loads. The pump drives for $\rm{JPC_1}$ and $\rm{JPC_2}$ are fed to the MPIJDA through two auxiliary ports denoted $\rm{P_1}$ and $\rm{P_2}$. The pump drives have the same frequency $f_p$, but can have different phases $\varphi_1$ and $\varphi_2$. As we show below, for a certain pump phase difference between the two pump drives, weak signals at frequency $f_1$ input on port $1$ of the MPIJDA are amplified upon exiting via port $2$. The JPCs are fabricated on high-resistivity $350$ $\mu\operatorname{m}$ thick silicon substrates. All conducting elements of the JPC are made of niobium, except the JRM which is made of aluminum and the ground plane on the back side of the substrate which is made of silver. The two JPC chips employed in the device have identical layout and are fabricated together on the same silicon wafer.

\section{The device theory}
To calculate the MPIJDA scattering parameters as well as to demonstrate its directionality, we use the effective signal-flow graph for the MPIJDA, exhibited in Fig. \ref{SignalFlow} (a). For simplicity, the various reflection and transmission parameters, indicated in the graph, correspond to on-resonance signals $f_1=f_a$ and $f_2=f_b$. As we show below, it is straightforward to generalize the device response for signals that lie within the JPC dynamical bandwidth $B$. We also assume that the two JPCs are fully balanced, i.e., their scattering parameters have the same amplitudes. Figure \ref{SignalFlow} (a) includes signal-flow graphs for JPCs operated in amplification mode. Signals input on port \textit{a} ($1^{\prime}$ or $2^{\prime}$) and \textit{b} (\textit{b1} or \textit{b2}) are reflected off with amplitude-gain $r$ and transmitted to the other port with amplitude-gain $s$, where $r$ and $s$ satisfy the condition $r^2-s^2=1$. The transmitted signals between ports $a$ and $b$ undergo frequency conversion and phase conjugation. The latter operation is represented in the signal-flow graph using unfilled arrows. The nonreciprocal phases $\varphi_1$ and $\varphi_2$, acquired by frequency-converted-transmitted signals between ports \textit{a} and \textit{b}, correspond to the phases of the pump drives at frequency $f_p$ feeding $\rm{JPC_1}$ and $\rm{JPC_2}$, respectively.    
In addition, Fig. \ref{SignalFlow} (a) includes flow graphs for two couplers coupling the signal and idler ports of the JPCs; one represents the $90^{\circ}$ hybrid, which couples between the signal ports of the JPCs, while the other is a fictitious one coupling the idler ports. The main role of the latter coupler is to model the amplitude attenuation present on the idler port $\alpha$, due to signal absorption in the $50$ Ohm cold loads terminating two feedlines of resonators \textit{b} and the insertion loss of the normal-metal transmission line coupling the two stages. Due to the structural symmetry of our device, we consider a generic, symmetric coupler with real coefficients $\alpha$ and $\beta$, which satisfy the condition ${\alpha}^2+{\beta}^2=1$. For an ideal symmetric coupler (i.e., $90^{\circ}$ hybrid), $\alpha=\beta=1/\sqrt{2}$ \cite{Pozar}. 

At resonance, and in the stiff pump approximation, the reflection and transmission gain amplitudes of the JPC can be written in the form \cite{JPCnaturePhys}
\begin{align}
\begin{array}
[c]{cc}%
r=\dfrac{1+\rho^2}{1-\rho^2}, \\
s=\dfrac{2\rho}{1-\rho^2},  
\end{array}
\label{r_s_res}%
\end{align}
where $0\leq\rho<1$ is a dimensionless pump amplitude. The lower bound $\rho=0$ corresponds to the case of no applied pump, whereas the limit $\rho\longrightarrow1^-$ corresponds to the case of high gain. By inspection \cite{Pozar}, the scattering matrix of the inner device defined by the ports $1^{\prime}, 2^{\prime}, 3, 4$, i.e., excluding the first $90^{\circ}$ hybrid, can be written in the form \cite{JDA}

\begin{align}
\left[  s\right]    & =\left(
\begin{array}
[c]{cccc}%
s_{1^{\prime}1^{\prime}} & s_{1^{\prime}2^{\prime}} & s_{1^{\prime}3} &
s_{1^{\prime}4}\\
s_{2^{\prime}1^{\prime}} & s_{2^{\prime}2^{\prime}} & s_{2^{\prime}3} &
s_{2^{\prime}4}\\
s_{31^{\prime}} & s_{32^{\prime}} & s_{33} & s_{34}\\
s_{41^{\prime}} & s_{42^{\prime}} & s_{43} & s_{44}%
\end{array}
\right)  \nonumber\\
& =\left(
\begin{array}
[c]{cccc}%
\frac{r\beta^{2}}{1-\alpha^{2}r^{2}} & \frac{\alpha s^{2}e^{-i\varphi}%
}{1-\alpha^{2}r^{2}} & \frac{\beta se^{-i\varphi_{1}}}{1-\alpha^{2}r^{2}} &
\frac{\beta r\alpha se^{-i\varphi_{1}}}{1-\alpha^{2}r^{2}}\\
\frac{\alpha s^{2}e^{i\varphi}}{1-\alpha^{2}r^{2}} & \frac{r\beta^{2}%
}{1-\alpha^{2}r^{2}} & \frac{\beta r\alpha se^{-i\varphi_{2}}}{1-\alpha
	^{2}r^{2}} & \frac{\beta se^{-i\varphi_{2}}}{1-\alpha^{2}r^{2}}\\
-\frac{\beta se^{i\varphi_{1}}}{1-\alpha^{2}r^{2}} & -\frac{\beta r\alpha
	se^{i\varphi_{2}}}{1-\alpha^{2}r^{2}} & -\frac{\beta^{2}r}{1-\alpha^{2}r^{2}}
& -\frac{\alpha s^{2}}{1-\alpha^{2}r^{2}}\\
-\frac{\beta r\alpha se^{i\varphi_{1}}}{1-\alpha^{2}r^{2}} & -\frac{\beta
	se^{i\varphi_{2}}}{1-\alpha^{2}r^{2}} & -\frac{\alpha s^{2}}{1-\alpha^{2}r^{2}}
& -\frac{\beta^{2}r}{1-\alpha^{2}r^{2}}%
\end{array}
\right),\label{s_inner_mat}%
\end{align}
where $\varphi\equiv\varphi_{1}-\varphi_{2}$. As we show below, it is this phase difference between the modulation phases of the two pumps feeding the two parametric active devices (i.e., the JPCs), which induces the nonreciprocal response of the MPIJDA. It generates an artificial gauge flux for photons \cite{NoiselessCirc,AhranovBohmPhotonic,AhranovBohmMixers,gyrator,DircJPC}, which plays the role of a magnetic field in a Faraday medium. The common coefficient $1/(1-\alpha^{2}r^{2})$ that appears in the scattering parameters of Eq. (\ref{s_inner_mat}) represents the sum over all possible reflections, that the internal signals can experience, in the self-loop formed between the two idler ports of the device. To ensure stability of the amplifier, the reflection-gain amplitude $r$ is bounded within the range $1\leq r<\alpha^{-1}$. In this simplified model, we assume that the phase acquired by signals at frequency $f_2$, propagating along the short transmission line between the two JPCs is $2\pi k$ in each direction, where $k$ is an integer. In our device, the electrical length of the short transmission line is designed to give a phase of about $2\pi$ at $f_2$.  

It is straightforward to verify that the scattering matrix in Eq. (\ref{s_inner_mat}) is symplectic (preserves information). For example, it satisfies the condition

\begin{equation}
\left\vert s_{1^{\prime}1^{\prime}}\right\vert ^{2}+\left\vert s_{1^{\prime
	}2^{\prime}}\right\vert ^{2}-\left\vert s_{1^{\prime}3}\right\vert
^{2}-\left\vert s_{1^{\prime}4}\right\vert ^{2}=1.
\end{equation} 

Next, we derive the scattering matrix for the whole device, defined by ports $1$, $2$, $3$, $4$, which take into account the signal flow through the $90^{\circ}$ hybrid, 

\begin{align}
\left[  S\right]    & =
\left(
\begin{array}
[c]{cccc}%
S_{11} & S_{12} & S_{13} & S_{14}\\
S_{21} & S_{22} & S_{23} & S_{24}\\
S_{31} & S_{32} & S_{33} & S_{34}\\
S_{41} & S_{42} & S_{43} & S_{44}%
\end{array}
\right),
\end{align}
whose matrix elements are given by,

%\begin{widetext}
\begin{align}
\left(
\begin{array}
[c]{cccc}%
\frac{1}{2}\left(  s_{1^{\prime}1^{\prime}}-s_{2^{\prime}2^{\prime}%
}+is_{2^{\prime}1^{\prime}}+is_{1^{\prime}2^{\prime}}\right)   & \frac{1}%
{2}\left(  is_{1^{\prime}1^{\prime}}+is_{2^{\prime}2^{\prime}}+s_{1^{\prime
	}2^{\prime}}-s_{2^{\prime}1^{\prime}}\right)   & \frac{1}{\sqrt{2}}\left(
is_{2^{\prime}3}+s_{1^{\prime}3}\right)   & \frac{1}{\sqrt{2}}\left(
is_{2^{\prime}4}+s_{1^{\prime}4}\right)  \\
\frac{1}{2}\left(  is_{1^{\prime}1^{\prime}}+is_{2^{\prime}2^{\prime}%
}+s_{2^{\prime}1^{\prime}}-s_{1^{\prime}2^{\prime}}\right)   & \frac{1}%
{2}\left(  s_{2^{\prime}2^{\prime}}-s_{1^{\prime}1^{\prime}}+is_{2^{\prime
	}1^{\prime}}+is_{1^{\prime}2^{\prime}}\right)   & \frac{1}{\sqrt{2}}\left(
s_{2^{\prime}3}+is_{1^{\prime}3}\right)   & \frac{1}{\sqrt{2}}\left(
s_{2^{\prime}4}+is_{1^{\prime}4}\right)  \\
\frac{1}{\sqrt{2}}\left(  is_{32^{\prime}}+s_{31^{\prime}}\right)   & \frac
{1}{\sqrt{2}}\left(  s_{32^{\prime}}+is_{31^{\prime}}\right)   & s_{33} &
s_{34}\\
\frac{1}{\sqrt{2}}\left(  is_{42^{\prime}}+s_{41^{\prime}}\right)   & \frac
{1}{\sqrt{2}}\left(  s_{42^{\prime}}+is_{41^{\prime}}\right)   & s_{43} &
s_{44}%
\end{array}
\right). 
\label{S_mat}%
\end{align}

%\end{widetext}

Note that since $[s]$ is symplectic and the $90^{\circ}$ hybrid is a unitary device, it directly follows that $[S]$ is symplectic as well. One prominent property of the device, as seen from Eq. (\ref{S_mat}), is its interferometric nature, which is manifested in the fact that the various scattering parameters of the device represent the sum over all possible paths that the waves can take in the device. 

By substituting the scattering parameters of the inner device listed in Eq. (\ref{s_inner_mat}) into Eq. (\ref{S_mat}), and by writing the resulting expressions in terms of the parameter $s$, we obtain the scattering parameters of the MPIJDA in an explicit form \cite{JDA}

\begin{align}
S_{21} &  =\frac{i}{1-\frac{\alpha^{2}}{\beta^{2}}s^{2}}\left[  \sqrt{1+s^{2}%
}+\frac{\alpha}{\beta^{2}}s^{2}\sin\varphi\right]  ,\label{S21}\\
S_{12} &  =\frac{i}{1-\frac{\alpha^{2}}{\beta^{2}}s^{2}}\left[  \sqrt{1+s^{2}%
}-\frac{\alpha}{\beta^{2}}s^{2}\sin\varphi\right]  ,\label{S12}\\
S_{11} &  =S_{22}=\frac{i\alpha}{\beta^{2}}\frac{s^{2}}{1-\frac{\alpha^{2}%
	}{\beta^{2}}s^{2}}\cos\varphi,\label{S11}\\
S_{33} &  =S_{44}=-\frac{\sqrt{1+s^{2}}}{1-\frac{\alpha^{2}}{\beta^{2}}s^{2}%
},\label{S33}\\
S_{34} &  =S_{43}=-\frac{\alpha}{\beta^{2}}\frac{s^{2}}{1-\frac{\alpha^{2}%
	}{\beta^{2}}s^{2}},\label{S34}\\
S_{13} &  =\frac{se^{-i\varphi_{s}  /2+i\pi/4}%
}{\sqrt{2}\beta\left(  1-\frac{\alpha^{2}}{\beta^{2}}s^{2}\right)  }\left[
\sqrt{1+s^{2}}\alpha e^{i\frac{\varphi}{2}+i\frac{\pi}{4}}+e^{-i\frac
	{\varphi}{2}-i\frac{\pi}{4}}\right]  ,\label{S13}\\
S_{14} &  =\frac{se^{-i  \varphi_{s}  /2+i\pi/4}%
}{\sqrt{2}\beta\left(  1-\frac{\alpha^{2}}{\beta^{2}}s^{2}\right)  }\left[ e^{i\frac{\varphi}{2}+i\frac{\pi}{4}}+\sqrt{1+s^{2}}\alpha e^{-i\frac
	{\varphi}{2}-i\frac{\pi}{4}}\right]  ,\label{S14}\\
S_{23} &  =\frac{se^{-i \varphi_{s}  /2+i\pi/4}%
}{\sqrt{2}\beta\left(  1-\frac{\alpha^{2}}{\beta^{2}}s^{2}\right)  }\left[
\sqrt{1+s^{2}}\alpha e^{i\frac{\varphi}{2}-i\frac{\pi}{4}}+e^{-i\frac
	{\varphi}{2}+i\frac{\pi}{4}}\right]  ,\label{S23}\\
S_{24} &  =\frac{se^{-i  \varphi_{s}  /2+i\pi/4}%
}{\sqrt{2}\beta\left(  1-\frac{\alpha^{2}}{\beta^{2}}s^{2}\right)  }\left[
e^{i\frac{\varphi}{2}-i\frac{\pi}{4}}+\sqrt{1+s^{2}}\alpha e^{-i\frac
	{\varphi}{2}+i\frac{\pi}{4}}\right]  ,\label{S24}\\
S_{31} &  =\frac{se^{i  \varphi_{s}  /2-i3\pi/4}%
}{\sqrt{2}\beta\left(  1-\frac{\alpha^{2}}{\beta^{2}}s^{2}\right)  }\left[
\sqrt{1+s^{2}}\alpha e^{-i\frac{\varphi}{2}+i\frac{\pi}{4}}%
+e^{i\frac{\varphi}{2}-i\frac{\pi}{4}}\right]  ,\label{S31}\\
S_{32} &  =\frac{se^{i  \varphi_{s}  /2-i3\pi/4}%
}{\sqrt{2}\beta\left(  1-\frac{\alpha^{2}}{\beta^{2}}s^{2}\right)  }\left[
\sqrt{1+s^{2}}\alpha e^{-i\frac{\varphi}{2}-i\frac{\pi}{4}}%
+e^{i\frac{\varphi}{2}+i\frac{\pi}{4}}\right]  ,\label{S32}\\
S_{41} &  =\frac{se^{i  \varphi_{s}  /2-i3\pi/4}%
}{\sqrt{2}\beta\left(  1-\frac{\alpha^{2}}{\beta^{2}}s^{2}\right)  }\left[
e^{-i\frac{\varphi}{2}+i\frac{\pi}{4}}+\sqrt{1+s^{2}}\alpha
e^{i\frac{\varphi}{2}-i\frac{\pi}{4}}\right]  ,\label{S41}\\
S_{42} &  =\frac{se^{i  \varphi_{s}  /2-i3\pi/4}%
}{\sqrt{2}\beta\left(  1-\frac{\alpha^{2}}{\beta^{2}}s^{2}\right)  }\left[
e^{-i\frac{\varphi}{2}-i\frac{\pi}{4}}+\sqrt{1+s^{2}}\alpha
e^{i\frac{\varphi}{2}+i\frac{\pi}{4}}\right]  ,\label{S42}
\end{align}
where $\varphi_{s}\equiv\varphi_{1}+\varphi_{2}$. In what follows, we examine a few special cases of interest and highlight a few important properties of the device. 

In the case of no applied pump, i.e., $s=0$, the MPIJDA scattering matrix reduces into

\begin{equation}
\left[  S\right]  =\left(
\begin{array}
[c]{cccc}%
0 & i & 0 & 0\\
i & 0 & 0 & 0\\
0 & 0 & -1 & 0\\
0 & 0 & 0 & -1
\end{array}
\right) \cdot\label{S_mat_s_zero}
\end{equation}

This result shows that when the device is \textit{off}, signals propagate between ports $1$ and $2$ with unity transmission in a reciprocal manner. In other words, in the \textit{off} state, the MPIJDA is transparent and mimics a lossless transmission line, which introduces a phase shift of about $\pi/2$ for signals within the bandwidth of the $90^{\circ}$ hybrid.
In the special case, where the MPIJDA is \textit{on} and the phase difference between the pumps is $\varphi=\pi/2$, the scattering matrix reads

\begin{equation}
\left[  S\right]  =\left(
\begin{array}
[c]{cccc}%
0 & i\left(  g-h\right)   & \sqrt{\frac{\left(  g-h\right)  ^{2}-1}{2}} &
\sqrt{\frac{\left(  g-h\right)  ^{2}-1}{2}}\\
i\left(  g+h\right)   & 0 & \sqrt{\frac{\left(  g+h\right)  ^{2}-1}{2}} &
\sqrt{\frac{\left(  g+h\right)  ^{2}-1}{2}}\\
-i\sqrt{\frac{\left(  g+h\right)  ^{2}-1}{2}} & -i\sqrt{\frac{\left(
		g-h\right)  ^{2}-1}{2}} & -g & -h\\
-i\sqrt{\frac{\left(  g+h\right)  ^{2}-1}{2}} & -i\sqrt{\frac{\left(
		g-h\right)  ^{2}-1}{2}} & -h & -g
\end{array}
\right)  ,\label{S_mat_symm_back_coup}%
\end{equation}
where 

\begin{align}
g &  =\frac{\sqrt{1+s^{2}}}{1-\frac{\alpha^{2}}{\beta^{2}}s^{2}},\label{gWalpha}\\
h &  =\frac{\alpha}{\beta^{2}}\frac{s^{2}}{1-\frac{\alpha^{2}}{\beta^{2}}s^{2}}.\label{hWalpha}%
\end{align}

In the derivation of Eq. (\ref{S_mat_symm_back_coup}), we assume, without loss of generality, that $\varphi_s=\pi/2$ (which corresponds to the case $\varphi_1=\pi/2$ and $\varphi_2=0$). As seen in Eq. (\ref{S_mat_symm_back_coup}), the amplifier, at this working point, is directional. The reflection parameters $S_{11}$ and $S_{22}$ vanish, whereas the amplitudes of the transmission parameters $S_{21}$ and $S_{12}$ are unequal. In the limit of high gain $s\rightarrow{(\beta/\alpha)}^{-}$, we obtain for the amplitudes of $S_{21}$ and $S_{12}$,

\begin{align}
\lim_{s\rightarrow{(\beta/\alpha)}^{-}}g+h  & =+\infty,\label{gPhsTo1}\\
\lim_{s\rightarrow{(\beta/\alpha)}^{-}}g-h  & =\frac{1+\alpha^2}{2\alpha},\label{gMhsTo1}%
\end{align}
respectively, where Eq. (\ref{gMhsTo1}) sets an upper bound on the reverse power gain of the device $|S_{12}|^2\leq(1+\alpha^2)^2/(4\alpha^2)$. In particular, for the case of a balanced coupler, i.e., $\alpha=\beta=1/\sqrt{2}$, this bound is $|S_{12}|^2\leq9/8$. Three important results of Eq. (\ref{gPhsTo1}) and Eq. (\ref{gMhsTo1}) are as follows: (1) Attaining arbitrary high gains in the forward direction is possible, (2) the backward gain of the device is bounded and it is on the order of unity, and (3) the forward gain is enabled by the self-loop in the system and the signal attenuation along this loop, i.e., partial routing of the internal mode towards auxiliary ports $3$ and $4$.   

If we further define 

\begin{align}
g+h &  =\sqrt{G},\label{sqrtG}\\
g-h &  =\sqrt{H},\label{sqrtH}%
\end{align}

we can cast Eq. (\ref{S_mat_symm_back_coup}) in an even simpler form%

\begin{equation}
\left[  S\right]  =\left(
\begin{array}
[c]{cccc}%
0 & i\sqrt{H} & \sqrt{\frac{H-1}{2}} & \sqrt{\frac{H-1}{2}}\\
i\sqrt{G} & 0 & \sqrt{\frac{G-1}{2}} & \sqrt{\frac{G-1}{2}}\\
-i\sqrt{\frac{G-1}{2}} & -i\sqrt{\frac{H-1}{2}} & -\frac{\sqrt{G}+\sqrt{H}}{2}
& -\frac{\sqrt{G}-\sqrt{H}}{2}\\
-i\sqrt{\frac{G-1}{2}} & -i\sqrt{\frac{H-1}{2}} & -\frac{\sqrt{G}-\sqrt{H}}{2} &
-\frac{\sqrt{G}+\sqrt{H}}{2}%
\end{array}
\right).  \label{S_mat_G_H}%
\end{equation}

In Fig. \ref{SignalFlow} (b), we exhibit a graphical representation of the direction and amplitude of the various scattering parameters of the device, based on the working point of Eq. (\ref{S_mat_G_H}). For clarity, we omit arrows representing amplitudes that are negligible. From Eq. (\ref{S_mat_G_H}) and Fig. \ref{SignalFlow} (b), it is straightforward to see that ports $3$ and $4$ play the role of auxiliary ports for the system that input vacuum noise, which is equivalent to the minimum amount of added noise required by quantum mechanics in the case of linear, phase-preserving amplification \cite{Caves,NoiseAmplReview}.     

Figure \ref{SignalFlow} (c) introduces a circuit symbol for the MPIJDA. For its operation, the device requires two pump drives at frequency $f_p$ having a phase difference $\varphi$. For $\varphi=\pi/2$, weak signals at frequency $f_1$ input on port $1$ of the device are transmitted with gain to port $2$ at the same frequency.  

Furthermore, to highlight the interferometric property of the MPIJDA, we illustrate in Fig. \ref{MultiPathInterference} the four principal paths that the waves take when signals propagate in the forward direction between ports $1$ and $2$. In these four paths, the signals enter port $1$ and exit through port $2$. In paths (a) and (b), the signal bounces off ports \textit{a} of $\rm{JPC_1}$ and $\rm{JPC_2}$, respectively, on its way to port 2. In paths (c) and (d), the signal is transmitted through $\rm{JPC_1}$ and $\rm{JPC_2}$ in (c) and $\rm{JPC_2}$ and $\rm{JPC_1}$ in (d). Large amplification is obtained in the forward direction when the waves propagating along these four principal paths constructively interfere. Similar principal paths can be drawn for signals propagating in the backward direction or reflecting off ports $1$ or $2$. In these directions, destructive interference taking place between the different paths eliminates or minimizes the amplitudes of the corresponding scattering parameters.    

It is worth noting that although the scattering parameters in Eqs. (\ref{S21})-(\ref{S42}) are derived for on-resonance signals, they can be generalized, in the vicinity of the resonance, by substituting 

\begin{align}
s[\omega_1]=\dfrac{2\rho}{\chi_{a}^{-1}\chi_{b}^{-1*}-\rho^2}, \label{s_param_vs_freq} 
\end{align}
where the $\chi's$ are the bare response functions of modes \textit{a} and \textit{b} (whose inverses depend linearly on the S and I frequencies): 

\begin{align}
\chi_{a}^{-1}[\omega_{1}]=1-2i\dfrac{\omega_{1}-\omega_{a}}{\gamma_{a}}, \label{Chi_a} \\ 
\chi_{b}^{-1}[\omega_{2}]=1-2i\dfrac{\omega_{2}-\omega_{b}}{\gamma_{b}}. 
\label{Chi_b}%
\end{align}

Since the angular frequency of the applied pump satisfies the relations $\omega_{p}=\omega_{a}+\omega_{b}=\omega_{1}+\omega_{2}$, Eq. (\ref{Chi_b}) can be rewritten in terms of $\omega_{1}$, 

\begin{align}
\chi_{b}^{-1}[\omega_{1}]=1+2i\dfrac{\omega_{1}-\omega_{a}}{\gamma_{b}}.
\label{Chi_b_mod}%
\end{align}

One important assumption of this generalization is that the $90^{\circ}$ hybrid has wider bandwidth than $\gamma_{a,b}$ of the JPCs, which is generally valid since transmission-line based hybrids have bandwidths on the order of several hundreds of megahertz \cite{CPWhybrids}. 

\begin{figure}
	[tb]
	\begin{center}
		\includegraphics[
		width=0.7\columnwidth 
		]%
		{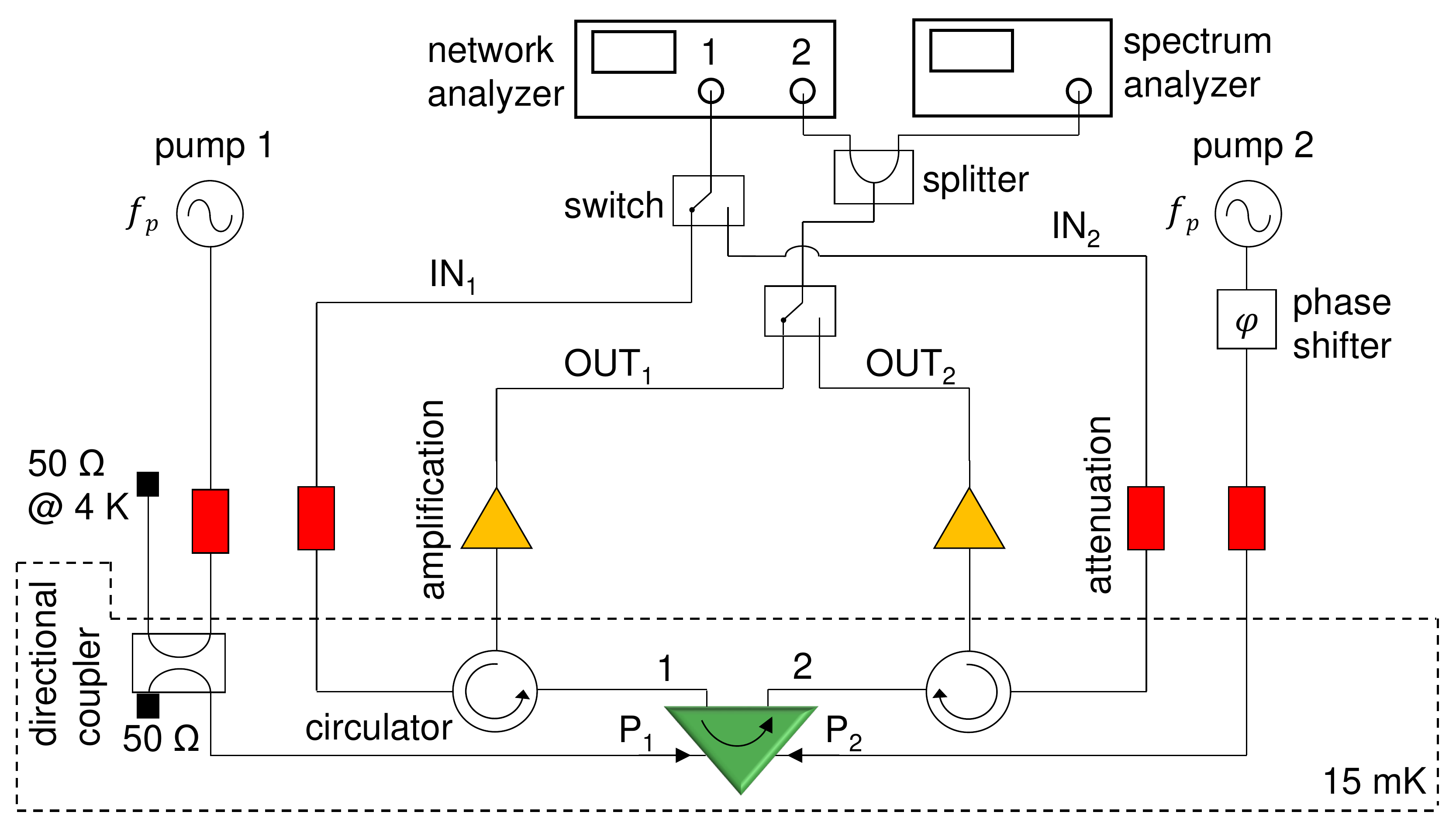}
		\caption{Block diagram of the experimental setup used in the measurement of the scattering parameters of the MPIJDA and its output spectrum. The diagram features only the main microwave components in the setup. It does not show the distribution of the attenuation (amplification) on the input (output) lines on the different temperature stages in the dilution fridge or the pair of isolators that are incorporated into each output line following the circulator. Each port of the device is connected to separate input and output lines via three-port circulators. The four scattering parameters of the device are measured using a network analyzer, whose two ports are connected to either one of the four possible pairs of input and output lines using switches located at room temperature. The setup also allows us to measure the output spectrum by connecting a spectrum analyzer, in parallel, to the network analyzer. The output signal is split between the two measuring devices using a power divider at room temperature. The pump drives are fed to the MPIJDA using two input lines. One pump line incorporates fixed attenuation at the $4$ $\operatorname{K}$ and mixing chamber stages, while the other incorporates fixed attenuation at the $4$ $\operatorname{K}$ stage and a directional coupler at the mixing chamber. The input pump power, in the latter line, is attenuated by the directional coupler. The unused portion of the pump power, in the latter line, is directed towards a $50$ Ohm termination at the $4$ $\operatorname{K}$ stage, whereas the reflected pump power off the device is dissipated in a $50$ Ohm termination at base. No observable heating of the mixing chamber has been observed in this experiment due to the application of the pump drives.  
		}
		\label{Setup}
	\end{center}
\end{figure}

\begin{figure}
	[tb]
	\begin{center}
		\includegraphics[
		width=0.8\columnwidth 
		]%
		{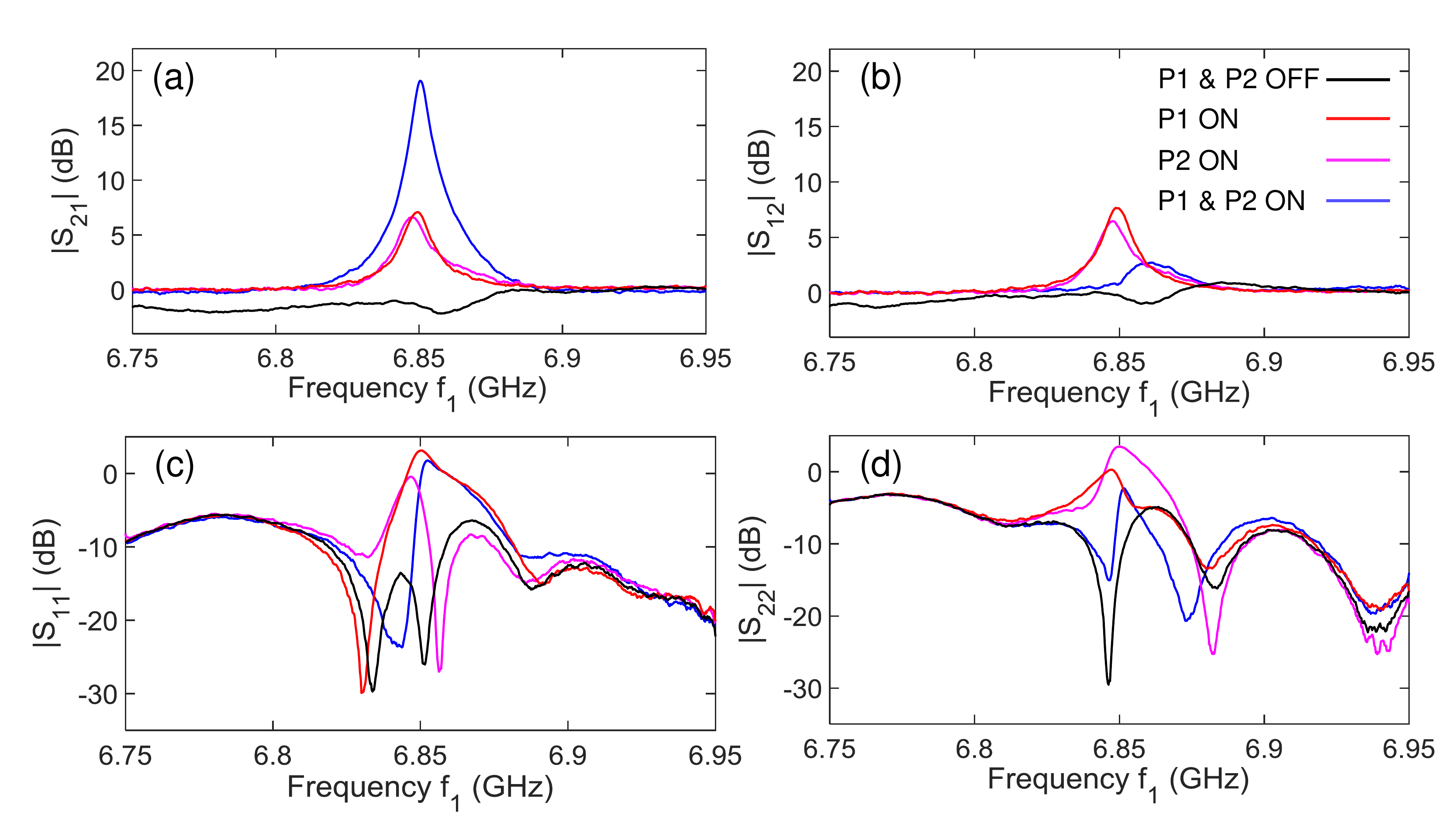}
		\caption{Scattering parameters measurement of the MPIJDA versus frequency taken at a directional amplification working point. Plots (a), (b), (c), and (d) exhibit measurements of forward ($|S_{21}|^2$), backward ($|S_{12}|^2$), input reflection ($|S_{11}|^2$), and output reflection ($|S_{22}|^2$) power gains, respectively. At this working point, the resonance frequency of $\rm{JPC_1}$ and $\rm{JPC_2}$ is flux-tuned to coincide at $6.85$ $\operatorname{GHz}$. The applied pump frequency is $16.599$ $\operatorname{GHz}$. The pump power feeding each JPC stage is set to yield a forward gain ($|S_{21}|^2$) of about $6.5$ dB with no pump applied to the other JPC. The pump phase difference between the two JPC stages is set to maximize the forward gain and minimize the reflection gains. The black and blue curves correspond to the device response when both pumps are \textit{off} and \textit{on}, respectively. The red and magenta curves correspond to the MPIJDA response when pump 1 (2) is \textit{on} (\textit{off}) and pump 2 (1) is \textit{on} (\textit{off}), respectively.     
		}
		\label{GainVsFreq}
	\end{center}
\end{figure}
\begin{figure}
	[tb]
	\begin{center}
		\includegraphics[
		width=0.8\columnwidth 
		]%
		{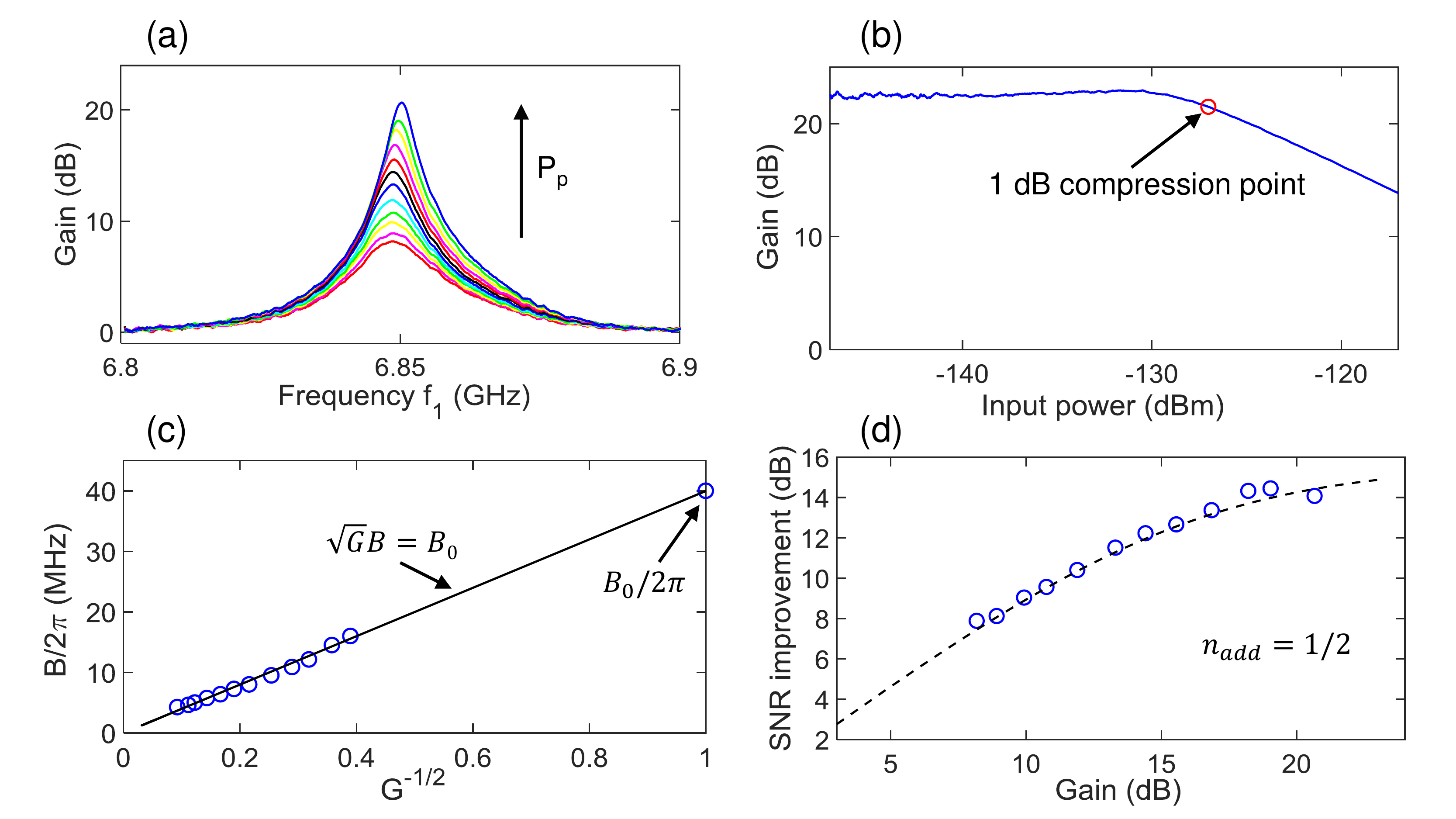}
		\caption{(a) Forward gain curves versus signal frequency $f_1$ measured at a fixed flux-bias point. The different colored gain curves correspond to different pump powers. (b) Maximum input power measurement taken on resonance at forward gain of $22.5$ dB. The red open circle marks the $1$ dB compression point of the MPIJDA. (c) Extracted dynamical bandwidth $B/2\pi$ of the signal (blue circles) versus $G^{-1/2}$. The solid black line represents the amplitude-gain bandwidth product relation. (d) SNR improvement measurement (blue circles) taken on resonance at port $2$ versus gain. The black dashed line corresponds to the calculated SNR improvement of the amplifier versus gain. The fit parameter $n_{\mathrm{add}}=1/2$ indicates that the MPIJDA operates near the quantum limit.  
		}
		\label{QLGvsBWDR}
	\end{center}
\end{figure}

\section{Experimental results}

\begin{figure}
	[tb]
	\begin{center}
		\includegraphics[
		width=0.8\columnwidth 
		]%
		{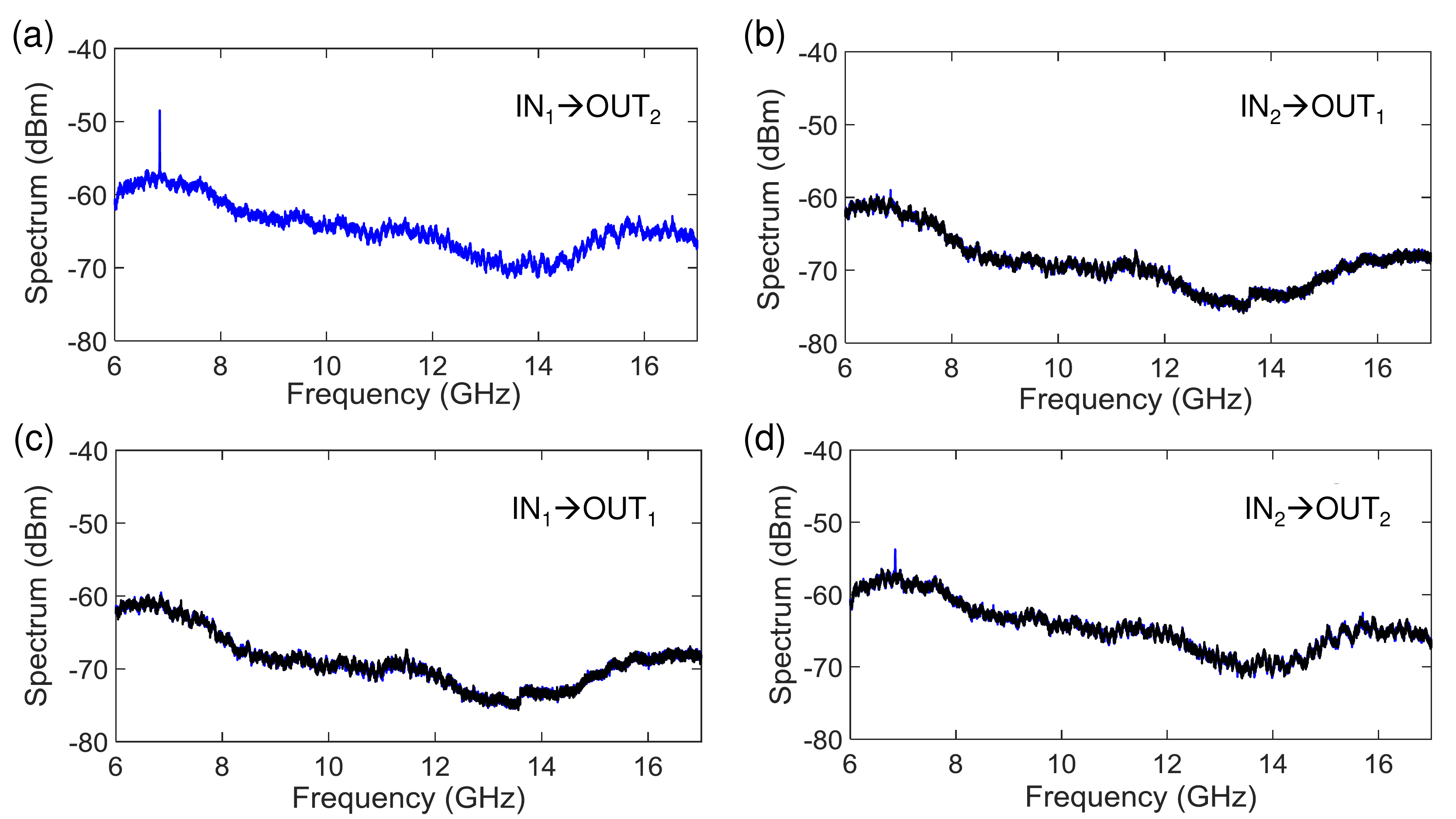}
		\caption{Plots (a), (b), (c), (d) exhibit the output spectrum of the MPIJDA measured at the directional amplification working point of Fig. \ref{GainVsFreq} for different input and output port combinations, as indicated by the labels in the panels. These measurements are taken using a spectrum analyzer and show the MPIJDA response in a wideband frequency range $6-17$ $\operatorname{GHz}$. The black and blue curves in each panel correspond to the pumps being \textit{off} and \textit{on}, respectively. In all these measurements, a $6.85$ $\operatorname{GHz}$ weak tone of about $-139$ dBm is applied to the input of the MPIJDA. No spurious harmonics or leakage of pump and idler signals are observed at the device ports. 
		}
		\label{Spectrum}
	\end{center}
\end{figure}

\begin{figure}
	[tb]
	\begin{center}
		\includegraphics[
		width=0.7\columnwidth 
		]%
		{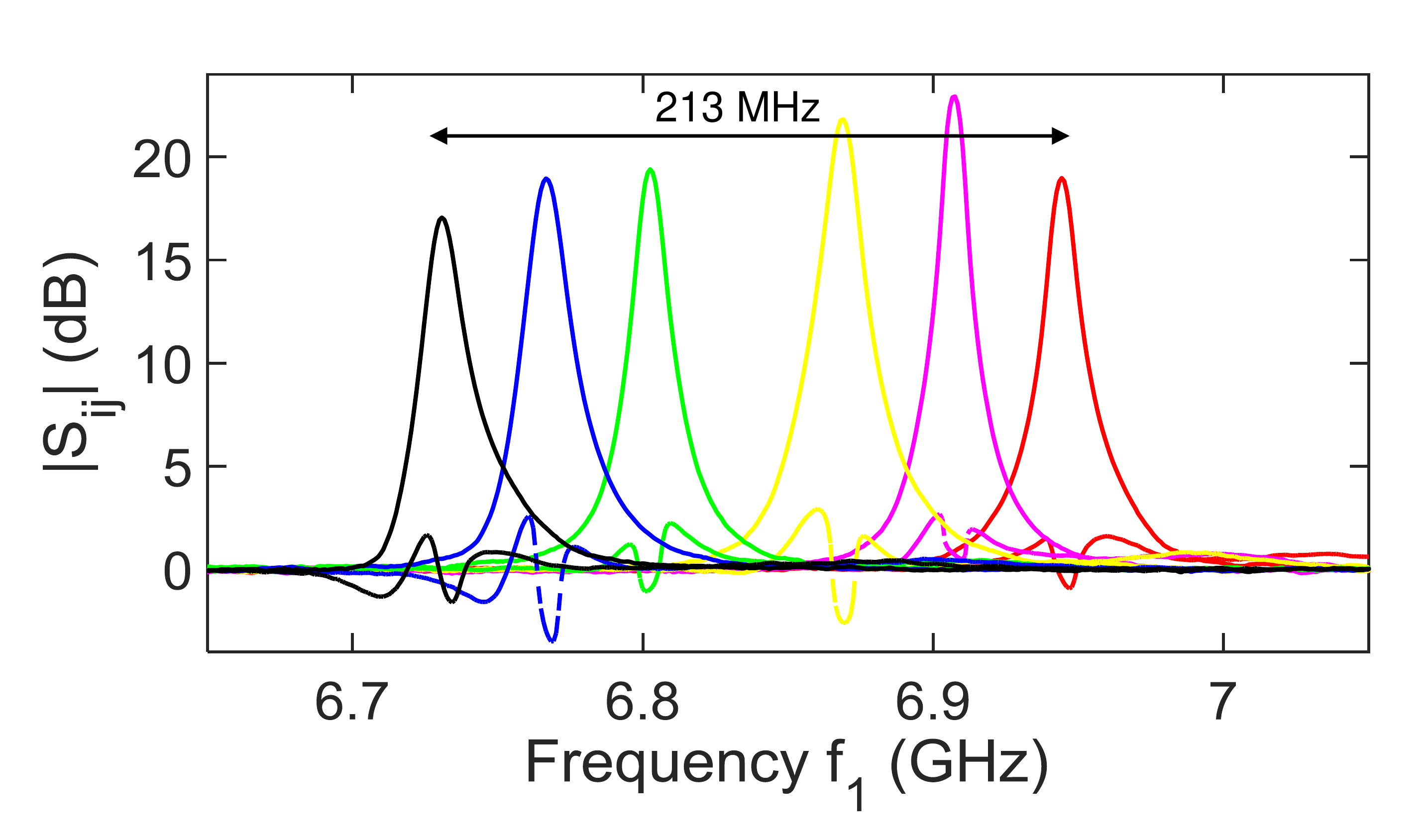}
		\caption{Tunable bandwidth measurement. The device forward (backward) gain $|S_{21}|^2$ ($|S_{12}|^2$) curves are measured for various flux points and drawn using solid (dashed) lines. The pump frequency and power are adjusted to yield a forward gain in excess of $17$ dB at each flux-bias point. The measurement shows a tunable bandwidth of about $213$ $\operatorname{MHz}$.   
		}
		\label{TunableBW}
	\end{center}
\end{figure}

\begin{figure}
	[tb]
	\begin{center}
		\includegraphics[
		width=0.8\columnwidth 
		]%
		{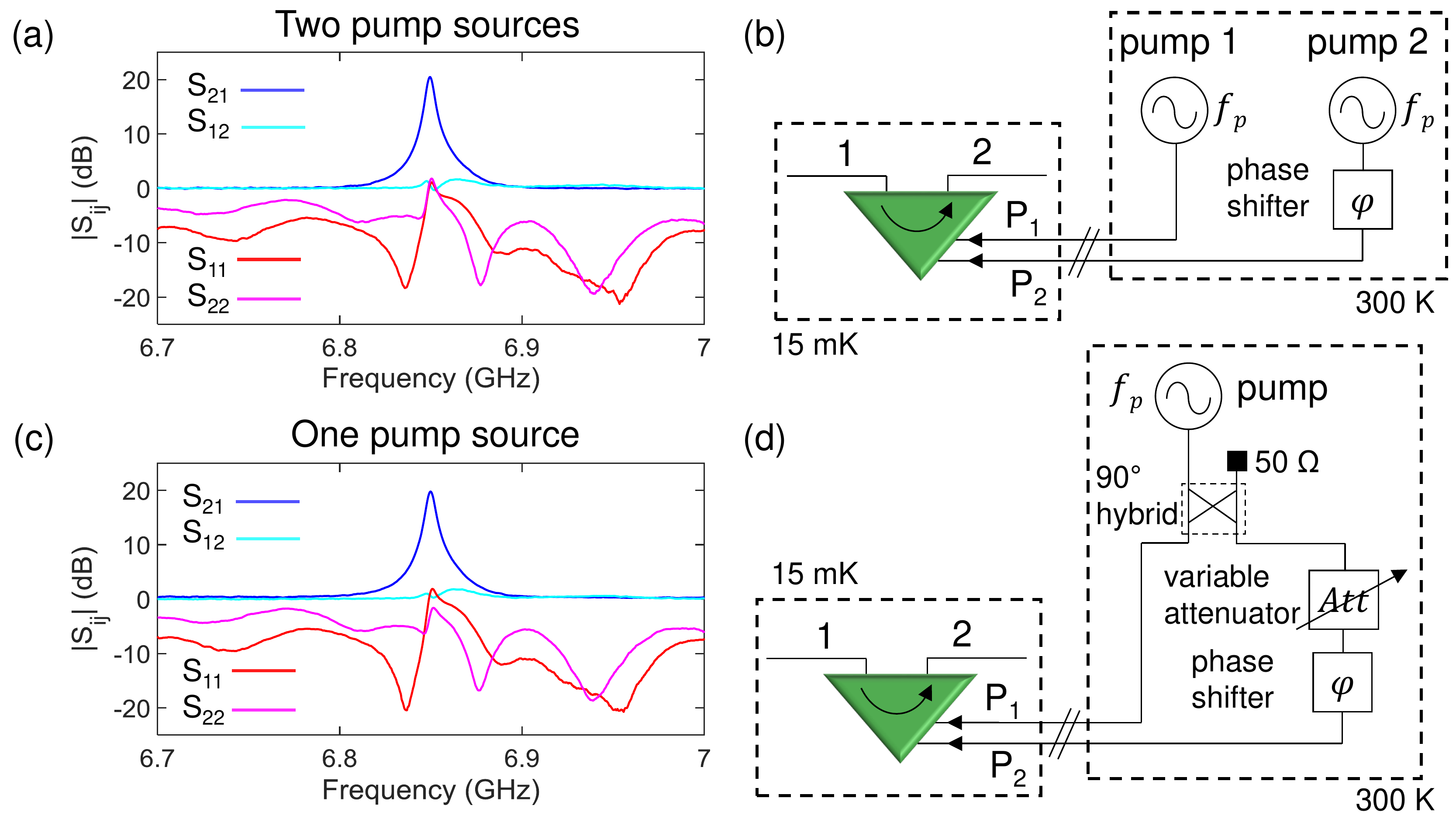}
		\caption{(a) and (c) Measured scattering parameters of the MPIJDA, i.e., $|S_{21}|^2$ (blue), $|S_{12}|^2$ (cyan), $|S_{11}|^2$ (red), $|S_{22}|^2$ (magenta), exhibiting directional amplification in the forward direction. The two pump drives feeding the MPIJDA in (a) are generated by two phase-locked microwave sources, versus one in (c) as illustrated in the setup diagrams shown in (b) and (d), respectively. In the case of the single microwave source, the pump drive is split using a $90^{\circ}$ hybrid into two arms, which connect to the pump input lines feeding the MPIJDA. A variable attenuator and a phase shifter is incorporated into one arm, at room temperature, which compensates for the attenuation difference between the two pump input lines and introduces the required pump phase shift between the two drives, respectively.   
		}
		\label{OneVsTwoP}
	\end{center}
\end{figure}

Figure \ref{Setup} exhibits a block diagram of the experimental setup used in the characterization of the MPIJDA. The scattering parameters of the device are measured using a network analyzer. Ports $1$ and $2$ of the MPIJDA are connected through circulators to separate input and output lines. Room temperature switches are used on ports $1$ and $2$ of the network analyzer to facilitate quick measurements of the four scattering parameters of the device. In this setup, port $1$ (transmitter) and $2$ (receiver) of the network analyzer are connected to the input and output lines of the MPIJDA, respectively. Some portion of the output signal coming out of the fridge is fed to a spectrum analyzer connected, in parallel, to the network analyzer via a power splitter. Pump drives $\rm{P_1}$ and $\rm{P_2}$ at frequency $f_p$, generated by two microwave sources, are fed to the device using two separate input lines. A phase shifter is used on one pump line to set the pump phase difference between the two drives. All microwave sources and measuring devices are phase locked to a $10$ $\operatorname{MHz}$ reference oscillator of a rubidium atomic clock.

In Fig. \ref{GainVsFreq}, we show measurements of the scattering parameters of the MPIJDA versus frequency, taken with the network analyzer. Plots (a) and (b) exhibit the transmission gain in the forward ($|S_{21}|^2$) and backward ($|S_{12}|^2$) directions, respectively. While plots (c) and (d) exhibit the reflection parameters off ports $1$ ($|S_{11}|^2$) and $2$ ($|S_{22}|^2$), respectively. The different colored curves in each plot correspond to different configurations of the pump drives. The various configurations are outlined in the legend of plot (b). The black curves correspond to the measured device response when both pumps $\rm{P_1}$ and $\rm{P_2}$ are \textit{off}. The red and magenta curves correspond to the measured device response when only one pump is \textit{on}, P1 and P2, respectively. Whereas the blue curves correspond to the MPIJDA response when both pumps are \textit{on} and a phase difference of $\pi/2$ is applied between them. It is worth noting that, prior to taking these measurements, the resonance frequencies of the two JPCs are aligned at about $6.85$ $\operatorname{GHz}$ by varying the external flux threading the two JRMs. As expected, when the device is \textit{off}, the black curves show that the reflections off the device ports are negligible (below $-20$ dB on resonance), whereas the transmission parameters are near $0$ dB. This result is in agreement with the prediction of Eq. (\ref{S_mat_s_zero}). For the measurements in which one pump drive is applied to either $\rm{JPC_1}$ or $\rm{JPC_2}$, i.e., the red and magenta curves, the pump power, in each case, is set to yield transmission gains of about $6.5$ dB in both directions ($|S_{21}|^2$ and $|S_{12}|^2$). Finally, for the measurements represented by the blue curves, in which both pumps are \textit{on}, the applied pump powers are identical to the ones set in the one pump measurements (red and magenta curves). In this measurement, the pump phase difference is set to maximize the forward gain, while minimizing the backward and reflection gains, as expected from Eq. (\ref{S_mat_symm_back_coup}). At the working point of Fig. \ref{GainVsFreq}, the MPIJDA yields a forward gain of $19.1$ dB, a backward gain of about $2.5$ dB (corresponding to a factor of $1.8$) and reflection gains of about $1.7$ dB for port $1$ and $-2.3$ dB for port $2$. Although, the reflection parameters of the device do not completely vanish, as expected from Eq. (\ref{S_mat_symm_back_coup}), this result shows that there exists a pump phase difference for which the backward and reflection gains are minimized while achieving a relatively high gain in the forward direction. In the following section, we show that these results can be largely attributed to having a nonideal $90^{\circ}$ hybrid connecting the JPCs to the device ports $1$ and $2$. We also discuss possible changes to the device design and implementation that could improve this figure of merit.  

Next, we measure several important characteristics of the amplifier. In Fig. \ref{QLGvsBWDR} (a), we measure the forward gain of the MPIJDA, versus frequency, for the same pump frequency and phase difference as Fig. \ref{GainVsFreq}, while varying the pump powers applied to the two JPCs. In this measurement, the pump powers are varied, in tandem, using the same power step. As seen in Fig. \ref{QLGvsBWDR} (a), the measured gain curves are Lorentzian and the maximum gain can be tuned smoothly and continuously from low to high values. In Fig. \ref{QLGvsBWDR} (b), we measure the maximum input power that the device can handle on resonance at $22.5$ dB of gain before it saturates, i.e., the input power for which the gain drops by $1$ dB. In this measurement, we find that the maximum input power of the device is about $-127$ dBm, which is comparable to maximum input powers achieved in microstrip JPCs \cite{JPCreview,Jamp}. In Fig. \ref{QLGvsBWDR} (c), we depict the dynamical bandwidths of the MPIJDA $B/2\pi$ (i.e., $-3$ dB points from the maximum gain) extracted from the measured gain curves in Fig. \ref{QLGvsBWDR} (a) versus $G^{-1/2}$. The blue circles correspond to the extracted $B$, while the solid black line corresponds to the amplitude-gain bandwidth product relation $B\sqrt{G}=B_{0}$, which is characteristic of Josephson parametric amplifiers employing resonators in the limit of large gains \cite{NoiseAmplReview,JPCreview}, where $B_{0}=2\gamma_{a}\gamma_{b}/(\gamma_{a}+\gamma_{b})$. The good agreement between the data and the amplitude-gain bandwidth product relation shows that, as expected, the MPIJDA bandwidth, similar to other JPAs, is bounded by this limit. 

To quantify the added noise by our phase-preserving amplifier, we measure the SNR improvement of the output chain, due to the MPIJDA, versus its gain $G$. The SNR improvement is given by $G/G_{N}=T_{N}/[T_{N}G^{-1}+T_{Q}(1/2+n_{\mathrm{add}})]$, where $T_{N}$ is the noise temperature of the output chain with the MPIJDA \textit{off}, $T_{Q}=hf_{2}/k_{B}$ is the effective temperature of an idler photon at frequency $f_{2}\cong9.75$ $\operatorname{GHz}$, and $n_{\mathrm{add}}$ is the added input photons by the amplifier. By fitting the data, we obtain $T_{N}=17$ $\operatorname{K}$, which agrees with our estimate of the noise temperature of the output line, and $n_{\mathrm{add}}=1/2$, which implies that the MPIJDA operates near the quantum limit in accordance with other JPCs \cite{JPCnature,JPCreview}.  

Furthermore, to verify that the MPIJDA, when operated at high forward-gain amplification points, does not generate undesired harmonics or exhibit power leakage between its ports, namely, P, $1$, and $2$, we measure, using a spectrum analyzer, the output spectrum of the MPIJDA in a wide frequency range $6-17$ $\operatorname{GHz}$, at the directional amplification working point of Fig. \ref{GainVsFreq}. In Fig. \ref{Spectrum}, the results are presented for the pumps \textit{off} (black) and \textit{on} (blue) cases. In all these measurements, a weak signal is applied to the MPIJDA input at $6.85$ $\operatorname{GHz}$. Plots (a), (b), (c), (d) exhibit the measured device spectrum for the (output, input) combinations (2,1), (1,2), (1,1), and (2,2). As seen in Fig. \ref{Spectrum}, the MPIJDA does not generate undesired harmonics or amplify noise away from resonance. It also lacks any observable power leakage between the device ports and the P ports, or its internal mode \textit{b}, at around $10$ $\operatorname{GHz}$.  

To measure the tunable bandwidth of the device, we vary the magnetic flux threading the two JRMs using the external, superconducting coils attached to the device package. By varying the external magnetic flux, the effective inductance of the JRM changes, which, in turn, shifts the resonance frequencies $f_a$ and $f_b$ of the JPC. In Fig. \ref{TunableBW}, we exhibit a tunable bandwidth measurement of the MPIJDA, in which we tune the resonance frequencies $f_{a1}$ and $f_{a2}$ together in steps of about $50$ $\operatorname{MHz}$. For each tuned resonance frequency, we set the frequency, the amplitudes, and the phase difference between the pump drives such that they yield a large gain in the forward direction (solid curves) and a small gain in the backward direction (dashed curves), respectively. This measurement shows that the MPIJDA yields forward gains in excess of $17$ dB and backward gains below $2.8$ dB within a frequency range of about $213$ $\operatorname{MHz}$.    

In Fig. \ref{OneVsTwoP}, we demonstrate that the MPIJDA does not require the use of two microwave sources for its operation; one source is sufficient. In Fig. \ref{OneVsTwoP} (a), we show a measurement of the four scattering parameters of the MPIJDA, i.e., $|S_{21}|^2$ (blue), $|S_{12}|^2$ (cyan), $|S_{11}|^2$ (red), and $|S_{22}|^2$ (magenta), operated at a directional amplification working point. In this measurement, the pump ports of the MPIJDA are driven by two separate microwave sources, as illustrated in the setup diagram shown in Fig. \ref{OneVsTwoP} (b). The phase difference between the two pump drives is set using a phase shifter connected to the output of one of the sources. In Fig. \ref{OneVsTwoP} (c), we repeat the measurement using one source, as illustrated in the setup diagram shown in Fig. \ref{OneVsTwoP} (d). In the repeated measurement, the pump drive is split into two at the output of the microwave source using a $90^{\circ}$ hybrid. The split pump drives are fed to the two pump ports of the device. A variable attenuator and a phase shifter are incorporated into the path of one of the split pump signals in order to compensate for the attenuation difference between the two pump lines and set the required phase difference between the pumps feeding the device. As seen in Figs. \ref{OneVsTwoP} (a) and (c), the excellent agreement between the original and reproduced measurement results, taken using two pump sources versus one, shows that the MPIJDA can be operated with a single microwave source.  

\section{Discussion and outlook}

\begin{figure}
	[tb]
	\begin{center}
		\includegraphics[
		width=0.8\columnwidth 
		]%
		{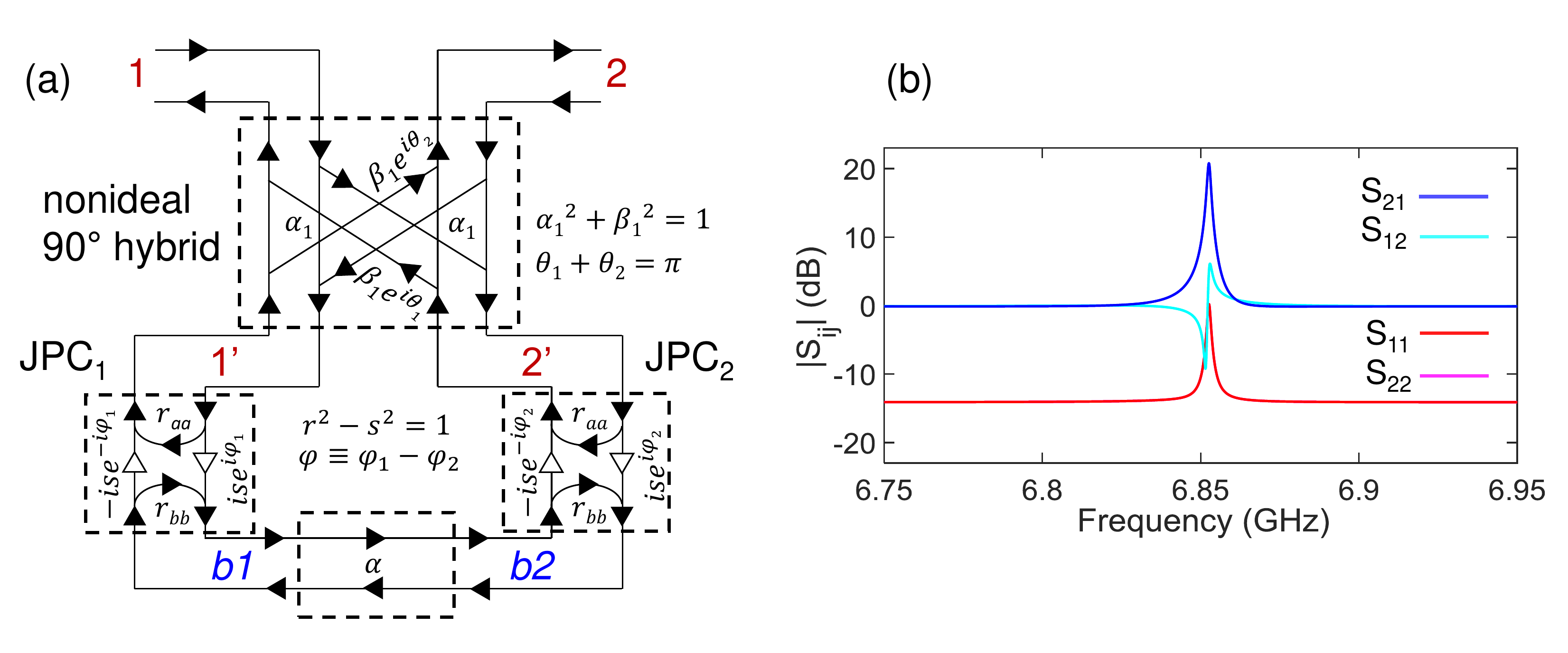}
		\caption{(a) Modified signal-flow graph for the MPIJDA. The modified graph employs a nonideal $90^{\circ}$ hybrid having a certain amplitude and phase imbalance between its arms and substitutes the 4-port, effective coupler connecting the \textit{b} ports of the two JPCs in Fig. \ref{SignalFlow} (a), with an attenuating transmission line with an attenuation amplitude $\alpha$. The nonideal hybrid has transmission amplitudes $\alpha_1$ and $\beta_1$, which satisfy the condition $\alpha_1^2+\beta_1^2=1$. The phase imbalance between the diagonal arms, represented by $\theta_1=\pi/2-\delta$ and $\theta_2=\pi/2+\delta$, satisfy the relation $\theta_1+\theta_2=\pi$. (b) Example of a calculated response of the MPIJDA using the modified signal-flow graph presented in plot (a). The calculation example reproduces the main features of the scattering parameters of the device, as seen in the measured response of Fig. \ref{OneVsTwoP} (a) and (c). The device and model parameters used in the calculation are: $\omega_a/2\pi=6.85$ $\operatorname{GHz}$, $\omega_b/2\pi=9.749$ $\operatorname{GHz}$, $\gamma_a/2\pi=\gamma_b/2\pi=40$ $\operatorname{MHz}$, $\varphi=\pi/2$, $|\alpha|=1/\sqrt2$, $\epsilon_{\rm{dB}}=0.4$ dB, $\delta=\pi/18$, and $\rho=0.38$.
		}
		\label{NonidealHybridSigFlowCalcS} 
	\end{center}
\end{figure}

Although, our present Josephson directional amplifier utilizes the same scheme and physics as the proof-of-principle device introduced in Ref. \cite{JDA}, it has several key enhancements, which make it particularly suitable for qubit readout in scalable architectures: (1) It uses hybridless JPCs, which allows the pump drives to be injected into the JPCs through separate physical ports from the \textit{a} and \textit{b} ports. Using this new configuration enables us to eliminate the four bulky, off-chip, broadband $180^{\circ}$ hybrids used in Ref. \cite{JDA} to address the \textit{a} and \textit{b} modes and also inject the pump drives. It also eliminates, as a byproduct, the eight phase-matched $2"$ coaxial lines that connect between the hybrids and the JPCs. 
(2) It realizes the $90^{\circ}$ hybrid and the transmission line connecting the two JPCs using a PCB. This new realization substitutes the off-chip, broadband $90^{\circ}$ hybrid and the $6"$ copper coax cable that connect the two idler feedlines in the proof-of-principle device. (3) It integrates the two JPCs into the same PCB and package. In the previous version, by contrast, the JPCs are mounted into two separate PCBs and housed inside two separate packages. Finally, (4) the new device resides in one cryoperm magnetic shield can, whereas the two JPCs of the previous device are placed in two separate cans. As a result of these enhancements, the footprint and weight of the present MPIJDA are significantly smaller compared to the proof-of-principle device. 

As seen in Figs. \ref{GainVsFreq}-\ref{OneVsTwoP} of the previous section, the experimental results of the MPIJDA are generally in good agreement with the JPC physics and the characteristics of the MPIJDA predicted by theory, namely, the gain-bandwidth product; the added noise; the nonreciprocal gain curves in the forward and backward directions; the tunable bandwidth; the maximum input power; the frequency spectrum; and the single pump operation of the device. One discrepancy, however, between the experimental results and the device theory is the measured reflection parameters of the device when the MPIJDA is \textit{on}. According to the device theory, the reflection parameters on resonance should vanish [see Eqs. (\ref{S11}) and (\ref{S_mat_G_H})], whereas, in the experiment, we observe on-resonance peaks of the reflection parameters of about $0$ dB. In order to resolve this discrepancy, we use a modified version of the signal-flow graph of the device exhibited in Fig. \ref{NonidealHybridSigFlowCalcS} (a). In the modified version, we replace the ideal $90^{\circ}$ hybrid, employed in the original signal-flow graph in Fig. \ref{SignalFlow} (a) with a nonideal version \cite{Pozar}, in which the hybrid's arms have a certain amplitude and phase imbalance, as outlined in Fig. \ref{NonidealHybridSigFlowCalcS} (a). We also focus our discussion on the four scattering parameters of the device, i.e., $S_{11}$, $S_{22}$, $S_{21}$, and $S_{12}$. To this end, we replace the effective coupler shown in Fig. \ref{SignalFlow} (a) with an attenuating transmission line between the idler ports with a transmission parameter $\alpha$. For the nonideal hybrid, we consider different transmission amplitudes $\alpha_1$ and $\beta_1$ for signals passing through the vertical and diagonal arms of the hybrid, respectively, where $\alpha_1$ and $\beta_1$ deviate from the ideal value $1/\sqrt2$, but satisfy the energy conservation relation $\alpha_1^2+\beta_1^2=1$. More specifically, $\alpha_1$ and $\beta_1$ are given by $\alpha_1^2=10^{(-3.01+\epsilon_{\rm{dB}})/10}$ and $\beta_1^2=10^{(-3.01-\epsilon_{\rm{dB}})/10}$, where $\epsilon_{\rm{dB}}$ is the power imbalance in decibels. We also assume different acquired phases for signals passing through the diagonal arms of the hybrid, i.e., $\theta_1=\pi/2-\delta$ and $\theta_2=\pi/2+\delta$, where $\delta$ is the phase imbalance and the phases $\theta_{1,2}$ satisfy the relation $\theta_1+\theta_2=\pi$. By inspection, the scattering parameters of the inner device, whose ports are $1^{\prime}$ and $2^{\prime}$, can be written as a function of the signal angular frequency in the form       

\begin{align}
\begin{array}
[c]{cc}%
s_{1^{\prime}1^{\prime}}[\omega_{1}]=r_{aa}+\dfrac{s^2r_{bb}\alpha^2}{1-r^2_{bb}\alpha^2}, & s_{1^{\prime}2^{\prime}}[\omega_{1}]=\dfrac{s^2e^{-i\varphi}\alpha}{1-r^2_{bb}\alpha^2},  \\
s_{2^{\prime}2^{\prime}}[\omega_{1}]=r_{aa}+\dfrac{s^2r_{bb}\alpha^2}{1-r^2_{bb}\alpha^2}, & 
s_{2^{\prime}1^{\prime}}[\omega_{1}]=\dfrac{s^2e^{i\varphi}\alpha}{1-r^2_{bb}\alpha^2},  \\
\end{array}
\label{s_nonidealhyb}%
\end{align}
where $s$ is given by Eq. (\ref{s_param_vs_freq}), the reflection parameters $r_{aa}$ and $r_{bb}$ read \cite{JPCreview} 

\begin{align}
\begin{array}
[c]{cc}%
r_{aa}[\omega_1]=\dfrac{\chi_{a}^{-1*}\chi_{b}^{-1*}+\rho^2}{\chi_{a}^{-1}\chi_{b}^{-1*}-\rho^2}, & 
r_{bb}[\omega_1]=\dfrac{\chi_{a}^{-1}\chi_{b}^{-1}+\rho^2}{\chi_{a}^{-1}\chi_{b}^{-1*}-\rho^2}, \\
\end{array}
\label{r_aa_r_bb}%
\end{align}
and the transmission parameter $\alpha=|\alpha|e^{i\varphi_{d}}$ incorporates the amplitude attenuation coefficient $|\alpha|$ and the phase shift $\varphi_{d}=\omega_2\tau_d=(\omega_p-\omega_1)\tau_d$ experienced by the I signals propagating between the two JPC stages at frequency $f_2$. In the expression for $\varphi_{d}$, $\tau_d$ represents the delay time, which can, in turn, be expressed as $\tau_d=l_d\sqrt{\varepsilon_d}/c$, where $c$ is the speed of light and $l_d=1.53$ $\operatorname{cm}$ is the electrical length of the stripline connecting the two stages.

By further using the relations of the nonideal $90^{\circ}$ hybrid, we arrive at the following expressions for the scattering parameters of the MPIJDA, as a function of frequency

\begin{align}
S_{11}[\omega_1] & =\alpha_1^2s_{1^{\prime}1^{\prime}}+\beta_1^2e^{2i\theta_1}s_{2^{\prime}2^{\prime}}+ 
\alpha_1\beta_1e^{i\theta_1}s_{2^{\prime}1^{\prime}}+\alpha_1\beta_1e^{i\theta_1}s_{1^{\prime}2^{\prime}}, \label{S_11_nonidealhyb} \\ 
S_{22}[\omega_1] & =\beta_1^2e^{2i\theta_2}s_{1^{\prime}1^{\prime}}+\alpha_1^2s_{2^{\prime}2^{\prime}}+
\alpha_1\beta_1e^{i\theta_2}s_{2^{\prime}1^{\prime}}+\alpha_1\beta_1e^{i\theta_2}s_{1^{\prime}2^{\prime}}, \label{S_22_nonidealhyb} \\ 
S_{21}[\omega_1] & =\alpha_1\beta_1e^{i\theta_2}s_{1^{\prime}1^{\prime}}+\alpha_1\beta_1e^{i\theta_2}s_{2^{\prime}2^{\prime}}+ 
\alpha_1^2s_{2^{\prime}1^{\prime}}-\beta_1^2s_{1^{\prime}2^{\prime}},  \label{S_21_nonidealhyb} \\ 
S_{12}[\omega_1] & =\alpha_1\beta_1e^{i\theta_2}s_{1^{\prime}1^{\prime}}+\alpha_1\beta_1e^{i\theta_2}s_{2^{\prime}2^{\prime}}- 
\beta_1^2s_{2^{\prime}1^{\prime}}+\alpha_1^2s_{1^{\prime}2^{\prime}}. \label{S_12_nonidealhyb} 
\end{align}

In Fig. \ref{NonidealHybridSigFlowCalcS} (b), we show a calculation result example for the scattering parameters of the MPIJDA with nonideal $90^{\circ}$ hybrid, where the red, magenta, blue, and cyan curves correspond to $|S_{11}|^2$, $|S_{22}|^2$, $|S_{21}|^2$, and $|S_{12}|^2$, respectively. In the calculation, we assume a power and phase imbalance of $\epsilon_{\rm{dB}}=0.4$ dB and $\delta=\pi/18$. As seen in Fig. \ref{NonidealHybridSigFlowCalcS} (b), the modified device model presented here successfully reproduces the main features of the scattering parameters of the device, in accordance with the measured response of Fig. \ref{OneVsTwoP} (a) and (c). In particular, it exhibits a significant nonreciprocal gain between the forward and backward directions and nonvanishing reflection gains near resonance of about $0$ dB.     

Based on these results, to further suppress the reflection gains of the device and potentially eliminate them, it is important to optimize the design and implementation of the $90^{\circ}$ hybrid employed in the MPIJDA scheme. Another indication that the present design and implementation of the $90^{\circ}$ hybrid is not optimal can be seen in the reflection parameters measured for the \textit{off} state of the MPIJDA, as displayed by the black curves in Fig. \ref{GainVsFreq} (c) and (d). In these figures, the magnitude of the reflection parameters do not exhibit a single large dip at resonance, with a relatively large bandwidth, which is expected for an ideal hybrid \cite{Pozar}. Also, the measured magnitude of the reflection parameters in the \textit{off} state, i.e., $|S_{11}|^2$ and $|S_{22}|^2$ exhibit different features. In contrast, in the case of an ideal hybrid, we expect the reflection parameters to be nominally identical.

In addition to optimizing the $90^{\circ}$ hybrid, an improved version of this amplifier could integrate all components on the same chip, i.e., the hybrid, the JPCs, and the cold terminations. An improved version could also replace the microstrip resonators of the JPCs, and possibly the hybrid, with lumped-element capacitors and inductors \cite{JPCreview,LumpedJPC,Lumpedhybrids}. Such a realization is expected to yield a miniature size that is mainly limited by the size of the external connectors or coils. 

Additional enhancements of the present design include the following: (1) Feeding the pump to the balanced JPCs through an on-chip or PCB-integrated $90^{\circ}$ hybrid centered around the pump frequency. In this configuration, the MPIJDA would be strictly driven by a single pump. The hybrid would equally split the pump drive into two parts that feed the JPCs and introduce a $\pi/2$ phase difference between them \cite{gyrator}. (2) Broadening the JPC bandwidth by engineering the impedance of the S and I feedlines. Similar impedance-engineering techniques applied to JPAs have successfully yielded large bandwidths of about $600$ $\operatorname{MHz}$ that break the gain-bandwidth product limit \cite{StrongEnvCoupling,JPAimpedanceEng}.         

\section{Conclusion}

We realize and characterize a two-port, phase-preserving, Josephson directional amplifier suitable for qubit readout. The device is formed by coupling two nominally-identical, hybridless JPCs through their signal and idler feedlines via a PCB-integrated $90^{\circ}$ hybrid and transmission line, respectively. Directional amplification is generated in the device using a novel reciprocity-breaking and reflection-minimizing mechanism that combines three processes: (1) generation of nonreciprocal phase shifts for signals transversing the JPCs in opposite directions, (2) amplification of signals using a self-loop and certain amount of attenuation for an internal mode of the system, and (3) formation of constructive and destructive wave-interference between multiple paths in the device.

On resonance, the directional amplifier exhibits a forward gain in excess of $20$ dB, a backward gain of about $2$ dB, and reflection gains on the order of $0$ dB. It has a dynamical bandwidth of about $7$ $\operatorname{MHz}$, a maximum input power on the order of $-127$ dBm, both at $20$ dB of gain, and a tunable bandwidth of about $200$ $\operatorname{MHz}$ with forward gains in excess of $17$ dB. 

Also, the directional amplifier, similar to single JPC devices, adds about a half input photon of noise to the processed signal, satisfies the amplitude-gain bandwidth product limit, does not generate undesired harmonics, preserves the signal frequency (analogous to the reflection gain in JPAs), and can be operated using a single microwave source.   

Looking forward, it is quite feasible to further suppress the reflection gains of the MPIJDA and enhance its dynamical bandwidth, by realizing optimized $90^{\circ}$ hybrids and applying certain impedance-engineering techniques to the feedlines of lumped-element JPCs, respectively. Furthermore, an improved version of the present device can be operated, by design, using a single microwave source by incorporating an on-chip or PCB-integrated $90^{\circ}$ hybrid for the pump, which delivers equal-amplitude, $\pi/2$ phase-shifted drives to the designated pump ports of the JPCs.

%\section*{Acknowledgments}
\begin{acknowledgments}
B.A. thanks Michel Devoret for interesting and fruitful discussions.  
\end{acknowledgments}

\end{document}